\documentclass{SciPost}

\hypersetup{
    colorlinks,
    linkcolor={red!50!black},
    citecolor={blue!50!black},
    urlcolor={blue!80!black}
}

\usepackage{bbm}
\usepackage[bitstream-charter]{mathdesign}
\usepackage{braket}

\DeclareSymbolFont{usualmathcal}{OMS}{cmsy}{m}{n}
\DeclareSymbolFontAlphabet{\mathcal}{usualmathcal}

\fancypagestyle{SPstyle}{
\fancyhf{}
\lhead{\colorbox{scipostblue}{\bf \color{white} ~SciPost Physics }}
\rhead{{\bf \color{scipostdeepblue} ~Submission }}

\fancyfoot[C]{\textbf{\thepage}}
}

\newcommand{\Tr}{\mathrm{Tr}}
\renewcommand{\Im}{\,\mathrm{Im}}

\def\({\left (}
\def\){\right )}

\begin{document}

\pagestyle{SPstyle}

\begin{center}{\Large \textbf{\color{scipostdeepblue}{
Out-of-equilibrium Eigenstate Thermalization Hypothesis \\ 
}}}\end{center}

\begin{center}\textbf{
Laura Foini\textsuperscript{1$\star$}
Anatoly Dymarsky\textsuperscript{2} 
Silvia Pappalardi\textsuperscript{3}
}\end{center}

\begin{center}
{\bf 1} IPhT, CNRS, CEA, Université Paris Saclay, 91191 Gif-sur-Yvette, France
\\
{\bf 2} Department of Physics and Astronomy, University of Kentucky, Lexington, KY 40503 USA
\\
{\bf 3} Institut für Theoretische Physik, Universität zu Köln, Zülpicher Straße 77, 50937 Köln, Germany
\\[\baselineskip]
$\star$ \href{mailto:email1}{\small laura.foini@ipht.fr}
\end{center}

\section*{\color{scipostdeepblue}{Abstract}}
\textbf{Understanding how out-of-equilibrium states thermalize under quantum unitary dynamics is an important problem in many-body physics. In this work, we propose a statistical ansatz for the matrix elements of non-equilibrium initial states in the energy eigenbasis, in order to describe such evolution. The approach is inspired by the Eigenstate Thermalisation Hypothesis (ETH) but the proposed ansatz exhibits different scaling.
Importantly, we point out the exponentially small cross-correlations between the observable and the initial state matrix elements that determine relaxation dynamics toward equilibrium.
We numerically verify scaling and cross-correlation, point out the emergent universality of the high-frequency behavior, and outline possible generalizations.
}

\vspace{\baselineskip}



\vspace{10pt}
\noindent\rule{\textwidth}{1pt}
\tableofcontents
\noindent\rule{\textwidth}{1pt}
\vspace{10pt}

\section{Introduction}

Over the past decades, the unitary evolution of nonequilibrium states, including post-quench dynamics, has been a prominent subject in the field of quantum dynamics.
The mechanism for thermalization 
is now well understood via the Eigenstate Thermalization Hypothesis (ETH) \cite{deutsch1991quantum, srednicki1999approach, Rigol_2008, dalessio2016from}. The latter is a statistical ansatz for the matrix elements of physical observables $\hat A$ is the energy eigenbasis $\hat H |E_i \rangle = E_i |E_i\rangle$:
\begin{equation}
	\label{ETH}
	A_{ij} = \mathcal A(E^+) \delta_{ij} +e^{-S(E^+)/2}f_A(E^+, \omega_{ij}) R_{ij}  \ ,
\end{equation}
with $E^+=(E_i+E_j)/2$, $\omega_{ij}=E_i-E_j$ being the average energy and frequency,  $S(E)$ is thermodynamic entropy, and $R_{ij}$ is a pseudorandom variable, such that $\overline{R_{ij}}=0$ and $\overline{R_{ij}R_{ji}}=1$. Finally, $\mathcal A(E)$ and $f_A(E, \omega)$ are smooth functions of their arguments. 
This ansatz has proved to be extremely successful in describing the equilibrium dynamics \cite{khatami2013fluctuation, dalessio2016from} of physical local Hamiltonians, as was shown by extensive numerical calculations \cite{prosen1999,rigol2008thermalization, biroli2010effect, ikeda2013finite, steinigeweg2013eigenstate, alba2015eigenstate, beugeling2015off, luitz2016long, fritzsch2021eigenstate}. Recently, the study of correlations between matrix elements \cite{FK2019} has led to novel developments beyond the standard framework, connecting ETH with Free Probability theory \cite{FK2019,Pappalardi:2022aaz, pappalardi2023general, fava2023designs}, random matrix universality \cite{ foini_RotInv, richter2020eigenstate, wang2021eigenstate, brenes2021out, foini_RotInv, jafferis2022matrix, jafferis2022jt,wang2023emergence}, conformal field theories \cite{deBoer2024multiboundary} and motivating the study of energy eigenvectors statistics \cite{Dymarsky:2016ntg,chan2019eigenstate,huang2019universal,PhysRevE.99.032111,murthy2019structure,shi2023local,hahn2023statistical, jindal2024generalized}. 

One of the central questions is how to extend the ETH framework to describe {\it non-equilibrium} dynamics 
\cite{srednicki1999approach,richter2019impact,dymarsky2019mechanism,knipschild2020modern,REIMANN2020121840,dymarsky2022bound,Capizzi:2024msk,deBoer2024}.
In this work, we propose a statistical ansatz for the matrix elements of the projector on the initial out-of-equilibrium state $\Psi=\ket{\psi}\bra{\psi}$ written in the eigenbasis of the Hamiltonian. 
Notably, the non-equilibrium dynamics are encoded in the \emph{correlations between the initial state and the observable's off-diagonal matrix elements}, which we describe in our framework.  
After introducing the ansatz and verifying its consistency, we discuss its implications for the relaxation dynamics towards equilibrium and numerically verify it in a non-integrable one-dimensional spin chain.
The main novelty of this work stems from the interplay between the standard ETH ansatz for the operator and the one that we propose for the projector via the correlations that characterize their matrix elements and that allows us to describe the non-equilibrium dynamics beyond the steady value.

\section{Out-of-equilibrium ETH}
\subsection{Set-up}

The dynamics of a local observable can be written in the eigenbasis of the Hamiltonian as
\begin{equation}
	\label{1}
	\langle \psi |\hat A(t) |\psi \rangle = \sum_{i j} c_i c_j^*\, A_{ij} e^{i(E_i - E_j)t} \ .
\end{equation}
with $c_i = \langle  \psi | E_i \rangle$.
The original ETH \eqref{ETH} is designed to describe the stationary equilibrium point. In the absence of degeneracies, the expectation value of $A$ eventually attains a stationary value
\begin{equation}
	\label{2}
	\sum_{i} |c_i|^2\, A_{ii} = \langle \hat A \rangle_{\rm diag}  \ ,
\end{equation}
which can be described by standard statistical mechanics. 
Namely, one introduces the diagonal ensemble $\hat \rho_{\rm diag}=\sum_i |c_i|^2 |E_i\rangle\langle E_i|$ such that $\langle \hat A \rangle_{\rm diag} = \text{Tr}\left ( \hat A \, \hat \rho_{\rm diag}\right )$ \cite{Rigol_2008, essler2016quench}.
In this work, we consider pure initial states with extensive mean energy and sub-extensive energy fluctuations in the number of degrees of freedom $N$ 
\begin{equation}\label{Scaling_energy}
\begin{array}{l}
\displaystyle
\langle \psi | \hat H | \psi \rangle = E_0 \simeq e_0 N,
\\ \vspace{-0.2cm} \\
\sqrt{\langle \psi | (\hat H- E_0)^2 | \psi \rangle} \simeq \delta_{e_0} N^{a}, \qquad a<1.
\displaystyle
\end{array}\ .
\end{equation}
 For such initial states, the stationary value of $\langle \hat A(t)\rangle$ is given by the microcanonical expectation, that, combined with ETH implies thermalization, i.e. $\langle \hat A \rangle_{\rm diag} \simeq \mathcal A(E_0)$ \cite{srednicki1999approach, Rigol_2008}.  
As a main example, we consider the case $a=1/2$, satisfied if one performs a global quench, which also characterizes equilibrium ensembles. Nonetheless, we will discuss the validity of our ansatz also for other initial states (see the Discussions).

The fundamental object that we want to characterize is the projector on the initial state written in the basis of the Hamiltonian
\begin{equation}
    \label{olap}
	\Psi _{ij}=  c_i c_j^{\ast} =  \langle E_j | \psi \rangle \langle \psi |E_i \rangle  \ .
\end{equation}
 We will treat it as a pseudorandom object, analogously to $A_{ij}$ in ETH.
A crucial difference, in comparison with  $\hat{A}$, is that this operator is of rank one and that each off-diagonal matrix element is the product of two pseudo-random numbers. This will radically change the scaling in the proposed ansatz \cite{PhysRevLett.122.070601}.
Crucially, to capture the out-of-equilibrium dynamics, we will assume that correlations exist between $\Psi$ and $\hat A$, when expressed in the energy eigenbasis.\\

\subsection{Ansatz}

We introduce an ansatz for the matrix $\Psi$,
\begin{equation}
    \label{eth_state}
 	\Psi _{ij} \simeq \frac {e^{- \Phi(E_i)}}Z \delta_{ij} + \frac{e^{-\frac 12 (\Phi(E_i)+\Phi(E_j))}}Z
  \,\tilde{R}_{ij} \ ,
\end{equation}
where $\Phi(E)$ is a smooth function of energy and  $Z=\sum_i e^{-\Phi(E_i)}$ the normalization, which defines the diagonal ensemble:
\begin{equation}
		\label{largeDV}
		\overline{\Psi_{ii}} = |c_{i}|^2 = \frac {e^{-\Phi(E_i)}}Z \ .
\end{equation}
The above average can be thought to be over small energy shells or over small perturbations, see e.g. \cite{deutsch1991quantum}.
In Eq.\eqref{eth_state}, $\tilde R_{ij}$ are pseudorandom variables with zero average and unit variance, i.e.
	\begin{equation*}
	\overline {\tilde R_{ij}} = 0 \ , \quad \overline {\tilde R_{ij}^2} = 1, \qquad \text{for $i\neq j$}.
    \end{equation*}
Diagonal pseudo-random $\tilde{R}_{ii}$ also have zero average, but the particular value of their variance may depend on the symmetry class of $\hat H$ (e.g.~GOE or GUE), as it is the case for  $R_{ii}$ in standard ETH \cite{mondaini2017,foini_RotInv}. Given that $\Psi_{ii}$ is positive, $(1+\tilde{R}_{ii}) \geq 0$.
Since $\Psi_{ij}$ is a product of two quasi-random numbers, this implies various constraints on the joint properties of $\tilde R_{ij}$. 
Crucially, these variables are exponentially weakly correlated with $R_{ij}$ of the original ETH ansatz \eqref{ETH}, 
 \begin{equation}
    \label{defg}
 	\overline{R_{ij}\tilde R_{ji}}= g_{A,\Psi}(e^+, \omega_{ij})\, e^{- S(e^+)/2} \ ,\quad  e^+=(E_i+E_j)/(2N),
 \end{equation}
where $ g_{A,\Psi}(e^+,\omega_{ij})$ is an order one smooth function of its variables, which describes the correlations crucial for non-equilibrium dynamics. Note that with our notations $e^+$ is the energy density between two eigenstates, while $e_0$, the energy density of the initial state in Eq.\eqref{Scaling_energy}. 
The existence of such correlations between the off-diagonal products of $\Psi_{ij} A_{ji}$, can be shown by averaging over the (random) phases of the eigenvectors of $\hat H$, see the App.\ref{app_sca}  Note that similar cross-correlations also exist between the matrix elements of different observables, say $\hat A$ and $\hat B$ \cite{noh202numerical, Pappalardi:2022aaz}. Within our notations this leads to a smooth function $F_{AB}^{(n)}(\omega)=f_A(\omega)f_{B}(-\omega)g_{AB}(\omega)$ with $g_{AB}(\omega_{ij}) = \overline{R_{ij}^A R^B_{ji}}$. .

 In particular, for the initial states conforming to  Eq.\eqref{Scaling_energy}, we will assume that a large deviation scaling of the form $\Phi(E) = N\phi(e=E/N)$ applies, such that the following is a \emph{convex function} \footnote{This implies $S''-\Phi''<0$, i.e. $\phi''(e)>s''(e)$.},
\begin{equation}
    \label{ext}
    S(E) - \Phi(E) = N [s(e) - \phi(e)]\ .
\end{equation} 	

Summarising, in the \emph{out-of-equilibrium ETH}, observables and the initial state look like pseudorandom  matrices with smooth statistical properties describing correlations or variance of the off-diagonal matrix elements,
\begin{subequations}
\label{outofeqETH}
\begin{align}
	\label{AA}
	\overline{|A_{ij}|^2} & = e^{-N s(e^+)} \, |f_A(e^+, \omega_{ij}) |^2, 
	\\ 	
	\label{BB}
	\overline{|\Psi_{ij}|^2 } & 
    = \frac{e^{-N(\phi(e_i)+\phi(e_j))}}Z^2
	\\
	\label{AB}
	\overline{\Psi_{ij} A_{ij}} & 
    = e^{-N s(e^+)} \,\frac {e^{- N \phi(e^+ )}}Z    f_A(e^+, \omega_{ij})\,  g_{A, \Psi}(e^+, \omega_{ij}),
\end{align}
\end{subequations}
where we made explicit the dependence on the system size $N$. 
In the case of states conforming to \eqref{Scaling_energy}, Eq.\eqref{BB} simplifies to
\begin{equation}
    \label{eq_BB_variance}
    \overline{|\Psi_{ij}|^2 }  
    = \frac {e^{- 2N \phi(e^+ )}}{Z^2} \,  e^{- \frac {\phi''(e^+)}{4 N}\omega_{ij}^2} \ ,
\end{equation} 
where one expands the energies $E_{i,j}=E^+\pm \omega_{ij}/2$ in Eq.\eqref{BB} around $E^+$ and 
uses the assumption of large deviation.

A similar term,  $e^{- \frac {\phi''(e^+)}{8N}\omega_{ij}^2}$, should appear also in Eq.\eqref{AB}, however, in the limit of large $N$ this can be neglected because $f_A(e, \omega) $ is expected to decay at large frequencies at least as $\exp(-\beta/4 \omega_{ij})$ \cite{murthy2019bounds} and only close-by eigenvectors contribute to quantum averages.\\
Finally, we note that the product $\Psi_{ij}A_{ij}$ has large fluctuations compared to the average \eqref{AB}, see the Appendix.\\

\subsection{Consistency checks}

Let us first see how the ansatz \eqref{eth_state} and in particular the diagonal ensemble derived from that, satisfy the assumptions (\ref{Scaling_energy}).
The large deviation scaling leads to an ensemble strongly peaked around the characteristic (extensive) energy which maximizes (\ref{ext}) and with sub-extensive fluctuations.
In fact, in the large $N$ limit, the energy uncertainty reads:
\begin{align}
 \label{Vari}
	\Delta^2_{E_0} \equiv \langle \psi | (\hat H-E_0)^2|\psi \rangle %
  =\frac{1}{\Phi''(E_0)-S''(E_0)} \ .
\end{align} 
Owing to the extensivity in Eq.\eqref{ext}, this implies that $\Delta_E$ is sub-extensive, in particular, for a post-quench state,
$$\Delta_{E_0} = \delta_{e_0} \sqrt{N}\ ,$$
where $\delta_{e_0}$ is an order-one constant, determined by the shape of the large deviation \footnote{Second derivative of an extensive quantity with respect to the energy  is $$
\Phi''(E_0) = \frac{\partial^2 \Phi(E)}{\partial E^2}\Big|_{E=E_0} = \frac 1N \frac{\partial^2 \phi(e)}{\partial e^2}\Big|_{e=e_0}= \frac 1N \phi''(e_0) .$$ Therefore $\delta_{e_0} = 1/\sqrt{\phi''(e_0)-s''(e_0)}$}.
We shall now proceed to discuss a set of consistency checks to validate our proposed approach.

\emph{Normalization --}
The state normalisation $\text{Tr}\, \Psi =1$
is ensured by the definition of $Z$. In the large $N$ limit,  the ``partition function" $Z$ in
Eq.\eqref{eth_state} reads:
\begin{align}
    \label{norma}
    Z & = {\sqrt{2\pi}}{\Delta_{E_0}} e^{S(E_0) - \Phi(E_0)} \ ,
\end{align}
with $\Delta_{E_0}$ given by Eq.\eqref{Vari}. However one can show a stronger property, namely that $\text{Tr}\, \Psi^2 = 1$. See Eq.\eqref{eq_gaussianFidelity} below at time zero.

\emph{Projector --} 
We now discuss an even tighter constraint: the projector identity $\Psi^2=\Psi$ at the level of individual matrix elements. For our ansatz, this turns out to be true in a statistical way in the thermodynamic limit.
In particular, thinking of the matrix elements $\Psi_{ij}$ as products of two random variables $\Psi_{ij}=c_i^{\ast} c_j$ we 
assume the following properties:
\begin{equation}\label{Eq_RRR}
\begin{array}{ll}
    \displaystyle
    \tilde{R}_{ik}\tilde{R}_{kj} = \tilde{R}_{ij}(1+\tilde{R}_{kk}), & \text{for $i\neq j \neq k$},
    \\ \vspace{-0.2cm} \\
    \displaystyle \tilde{R}_{ij}\tilde{R}_{ji} = (1+\tilde{R}_{ii}+\tilde{R}_{jj}+\tilde{R}_{ii}\tilde{R}_{jj}), & \text{for $i\neq j$}.
\end{array}
\end{equation}
At the leading order in $N$, this implies (see the Appendix),
\begin{subequations}\label{Eq_Psi2}
\begin{align}
[\Psi^2]_{ij} & \simeq \frac{e^{-\frac{1}{2}(\Phi(E_i)+\Phi(E_j))}}{Z} \tilde{R}_{ij} \simeq [\Psi]_{ij} \quad i\neq j,
\\
[\Psi^2]_{ii} & \simeq \frac{e^{-\Phi(E_i)}}{Z} \left( 1+ \tilde{R}_{ii}\right) \simeq [\Psi]_{ii}.
\end{align}
\end{subequations}
Therefore the ansatz preserves its structure upon multiplication.

\subsection{Statistical distribution of the matrix elements}
In the spirit of Berry conjecture \cite{berry1977regular}, or by analogy to Random Matrix Theory,  at leading order, one can expect individual coefficients $c_i$, rescaled by their typical value, to be Gaussian random numbers. For the distribution of the diagonal elements $\Psi_{ii}=|c_i|^2$, this leads to the well-known Porter-Thomas distribution for $x = \Psi_{ii}/\overline{\Psi_{ii}}$ \cite{porter1056fluctuation, haake1991quantum}. \\
We apply the same argument to the distribution of the re-scaled off-diagonal matrix elements $z = \Psi_{ij}/\sqrt{\overline{ \Psi_{ij}\Psi_{ji}}}=\tilde R_{ij}$, treating them  as a product of two independent  Gaussian  random variables.
In the thermodynamic limit, it yields the modified Bessel function
\begin{equation}
    \label{eq_bessel}
    P(z) = \frac 1{\pi}{K_0(|z|)} \simeq \frac{1}{\sqrt{2\pi}}\frac{e^{-|z|}}{\sqrt {|z|}} \ ,
\end{equation}
where on the right-end side, we substituted the asymptotic expansion for large $z$.\\

\subsubsection{Ansatz for the higher-order cross correlations}
Note that Eq.\eqref{eq_bessel} describes the distribution of the rescaled matrix elements $\tilde R_{ij}$, but it does not contradict the existence of their cross-correlations. This is analogous to the ETH matrix elements $R_{ij}=A_{ij}/\sqrt{\overline{A_{ij}A_{ji}}}$, which are known to  follow Gaussian distribution \cite{luitz2017anomalous, leblond2019entanglement, brenes2020low}, while, within RMT, the random variables $A_{ij}$ follow a large deviation function \cite{foini2022annealed,foini_RotInv} and
different matrix elements $A_{ij}$ are correlated \cite{FK2019,
dymarsky2022bound,wang2021eigenstate,foini_RotInv}.
In the case of  $\Psi_{ij}$,  in addition to \eqref{Eq_RRR}, there is an infinite number of relations coming from the fact that it is a projector, see Appendix \ref{App_norm}. 

Our main focus is on the correlations between $\Psi$ and the observables, such as the ones in Eq.~\eqref{AB}. For a multiple product with the distinct indices $i_1 \neq i_2 \neq i_n$, one can write
\begin{equation}
    \overline{\Psi_{i_1 i_2} A_{i_2 i_3} ... A_{i_n i_1}} = e^{- (n-1) S(e_+)} \frac{e^{- \Phi(E_+)}}Z
    h_{e^+}^{(n)}(\vec \omega)\ ,
\end{equation}
while contributions with the repeated indices factorize, i.e.
\begin{equation}
    \overline{\Psi_{i_1 i_2}  ...A_{i_{k-1} i_1} A_{i_{1} i_{k+1}} \dots  A_{i_n i_1}} =
    \overline{\Psi_{i_1 i_2}  ...A_{i_{k-1} i_1}}\cdot \overline{ A_{i_{1} i_{k+1}} \dots  A_{i_n i_1}}  + O({\mathcal D}^{-1})\ .
\end{equation}
Here $h_{e^+}^{(n)}(\vec \omega)$ is a smooth function of its arguments: $e^+=(E_{i_1}+\dots E_{i_n})/N n$  and
$\vec \omega=(E_{i_1}-E_{i_2}$ $, \dots, E_{i_n}-E_{i_{n-1}})$.
This ansatz encodes the out-of-equilibrium multi-time dynamics such as \\
$\bra{\psi} \hat A(t_1) \hat A(t_2) \dots \hat A(t_{n-1})\ket{\psi}$ and also ensures that the ansatz on the initial state is stable against perturbations.

\subsection{Implications for the dynamics}

We now discuss the main motivation behind our ansatz, designed to describe equilibration dynamics of physical observables. 

\emph{Fidelity decay --} First of all, we show that our ansatz is consistent with the expected behavior of the fidelity decay (survival probability) \cite{gorin2006dynamics, torres2014local, schiulaz2019thouless, rai2023matrix}, defined as
\begin{align}
\label{eq_fidelity}
	| \langle \psi |\psi(t)\rangle|^2 & = \left |\sum_i |c_i|^2 e^{-i E_i t} \right |^2\ .
 \end{align}
By substituting sums with integrals, neglecting spectral correlations, and using the out-of-equilibrium ETH ansatz in Eq.\eqref{eq_BB_variance}, at the leading order in $N$, one finds
\begin{align}
\begin{split}
| \langle \psi |\psi(t)\rangle|^2 
 \simeq \frac {1}{2\Delta_{E_0}\sqrt \pi} \int d\omega e^{- \frac 12 \frac 1{2 \Delta^2_{E_0}}  \omega^2}e^{i\omega t}
	  = e^{- t^2\,\Delta^2_{E_0} }\ ,
   \label{eq_gaussianFidelity}
\end{split}
\end{align}
where we used the definition of the energy variance in Eq.\eqref{Vari}, $\Delta_{E_0} =  1/ \sqrt{\Phi''(E_0) - S''(E_0) }$ $ = \delta_{e_0} \sqrt N$. 
Thus the large deviation ansatz in Eq.\eqref{eth_state} is consistent with the Gaussian decay, 
controlled by the energy variance of the initial state, in agreement with the literature on global quenches, see e.g. Ref.~\cite{torres2014local}.
Dynamical behavior, different from Eq.\eqref{eq_gaussianFidelity}, e.g.~including an exponential decay, is known to arise from the initial states which are different from Eq.\eqref{Scaling_energy} \cite{gorin2006dynamics, santos2012chaos, PhysRevE.85.036209, torres2014local, schiulaz2019thouless, micklitz2022emergence, long2023beyond}, as discussed below.

\emph{Relaxation dynamics --}
The primary purpose of this work is the study relaxation dynamics in Eq.\eqref{1}, namely:
\begin{equation}
	\label{eqrel}
	\delta A_{\Psi}(t) = \langle \psi |\hat A(t) |\psi\rangle  - \langle \hat A\rangle_{\rm diag} = \sum_{i\neq j}
	 c_i c_j^*\, A_{ij} e^{i(E_i - E_j)t}\ .
\end{equation}
Plugging Eq.\eqref{AB} into Eq.\eqref{eqrel}, the standard ETH manipulations lead to
\begin{align}
    \label{dAt}
	\delta A_{\Psi}(t) 
& \simeq \int d\omega \, f_A(e_0, \omega) g_{A, \Psi}(e_0,  \omega)  e^{-\frac{\omega^2}{4 N \delta^2_{e_0}}}e^{-i \omega t} \ .
\end{align}
Hence, the correlation between the initial state and the operator encodes the Fourier transform of the relaxation:
\begin{equation}
	\tilde{\delta A_{\Psi}}(\omega) = f_A(e_{0}, \omega) g_{A, \Psi}(e_{0},  \omega) \ ,
\end{equation}
where we have neglected the Gaussian frequency term for $N\gg 1$.
Thus, the out-of-equilibrium behavior is encoded in this function and will depend, in general, on the 
correlations between the state and the observable.

The relaxation dynamics (\ref{eqrel}) share some properties with the (two-time) dynamical correlations at thermal equilibrium for the same observable. 
One has \cite{khatami2013fluctuation, dalessio2016from}:
\begin{equation}\label{Equilibrium_dyn}
C(t) = \frac12 \langle \{ A(t), A(0) \}\rangle_c = \int \, {\rm d} \omega \, e^{i \omega t} \cosh \left (\frac{\beta\omega}{2} \right )f_A^2(e_{\beta}, \omega)\ ,
\end{equation}
where $\langle \cdot\rangle = \Tr(e^{-\beta H}\cdot )/\Tr(e^{-\beta H})$ and $e_\beta = \langle H \rangle/N$.
Therefore the ETH function $f_A(e, \omega)$ enters both Eqs.~\eqref{dAt} and \eqref{Equilibrium_dyn} and its properties in the $\omega \to 0$ limit control the long-time behavior. This fact is usually invoked in the literature, see e.g. \cite{srednicki1999approach,khatami2013fluctuation}, and our ansatz in Eq.~\eqref{dAt} makes it explicit.\\
Similarly to $f_A^2(\omega)$, which has to decay exponentially at large $\omega$ in $D\geq 2$ (superexponentially in 1D), high-frequency tail of $g_{\Psi,A}(\omega)$ has to be exponentially suppressed for states $\Psi$ associated with local perturbations (see the Appendix). 

Let us now comment on some differences between Eq.~\eqref{dAt} and \eqref{Equilibrium_dyn}. The integrand in  \eqref{Equilibrium_dyn} is positive-definite. As a result, $C(t)$ necessarily decays at early times. On the contrary, the integrand in Eq.~\eqref{dAt} is not sign-definite, hence  $\delta A_\Psi(t)$ can both increase or decrease throughout relaxation dynamics.

\section{Numerical results} 

 \begin{figure}[t]
\centering
\includegraphics[width=1\linewidth]{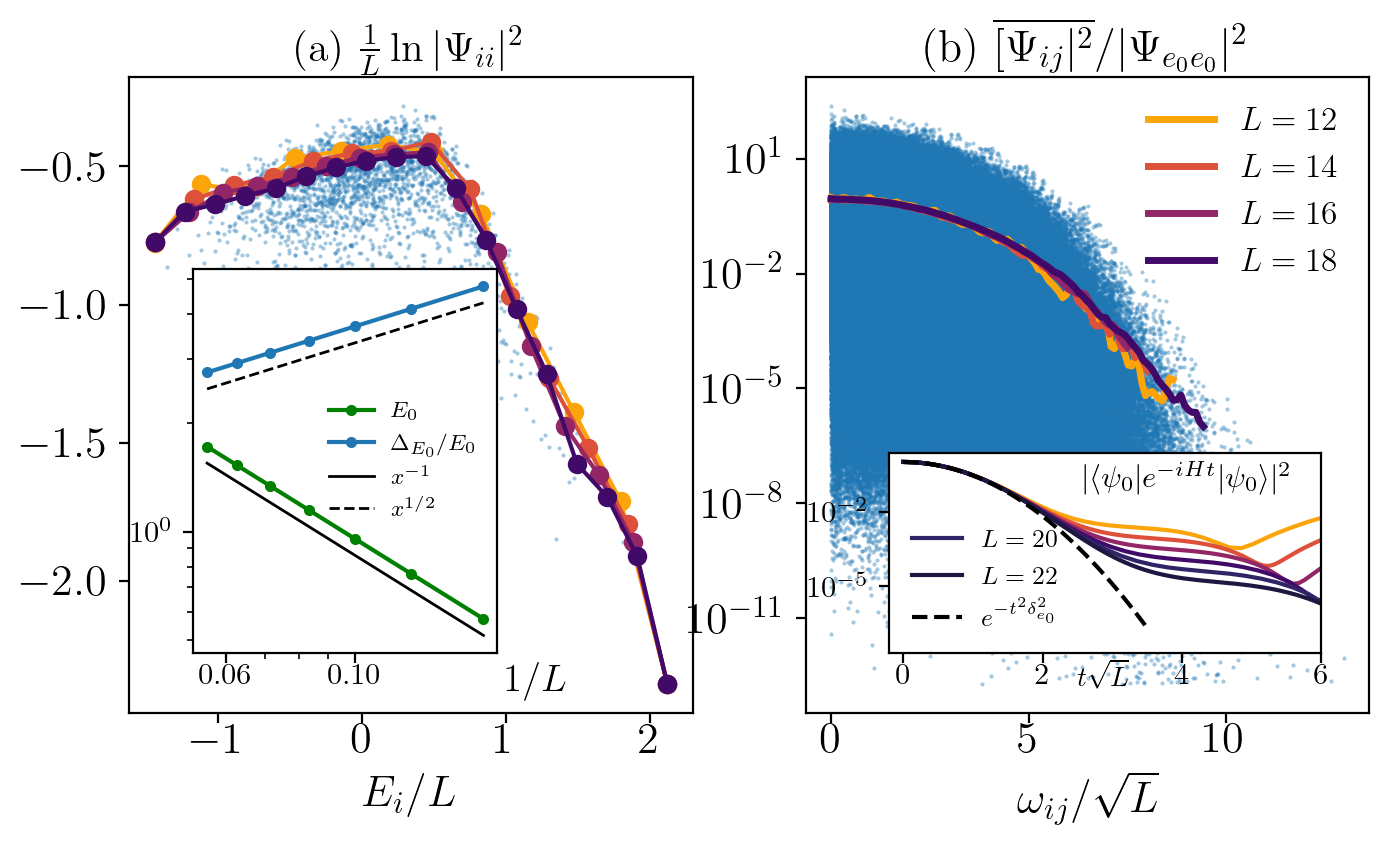}
\caption{Out-of-Equilibrium ETH of the fully polarized initial state ${|\psi\rangle=|\downarrow_z\dots \downarrow_z\rangle}$. (a) The rescaled diagonal ensemble as a function of the energy density for different system sizes $L=12,14,16,18$. In the inset, the initial energy $E_0=\langle \psi|\hat H|\psi\rangle$ and the energy fluctuations $(\Delta E_0)^2 = \langle \psi|\hat H^2|\psi\rangle-E_0^2$ are plotted as a function of the inverse of the system size $1/L$. 
 (b)  The off-diagonal matrix elements of the projector $\Psi_{ij}$ as a function of the energy difference $\omega_{ij}=E_i-E_j$ rescaled by $1/\sqrt{L}$.
In both panels, the blue dots correspond to individual overlaps for $L=16$. The smoothing parameter is $\tau=4$. In the inset of panel (b), the numerical fidelity decay up to $L=22$ is compared with the prediction of Eq.\eqref{eq_gaussianFidelity} (dashed line) without any fitting parameters. 
 }
\label{Fig1}
\end{figure}
We test the predictions above in the case of the one-dimensional Ising model with a tilted field
\begin{equation}
\label{H}
H = \sum_{i=1}^L w \sigma_i^x + \sum_{i=1}^L h \sigma_i^z + \sum_{i=1}^{L-1}J \sigma_i^z \sigma_{i+1}^z
\end{equation}
with $w=\sqrt{5}/2$, $h =  (\sqrt{5} + 5)/8$ and  $J=1$ and consider different local single or two sites observables,
\begin{equation}
    \label{obs}
    \hat A = \sigma^x_1, \, \hat A = \sigma^z_1 \, \quad \text{or}\quad \hat A = \sigma^z_1\sigma^z_{2} \ .
\end{equation}
We consider simple  out-of-equilibrium initial states, fully polarized states 
\begin{equation}
    \label{eq_psi0}
    |\psi\rangle = \ket{\downarrow_\alpha \downarrow_\alpha \dots \downarrow_\alpha }
\end{equation}
in the  $\alpha=z$ or $\alpha=y$ directions. 
We impose periodic boundary conditions on the Hamiltonian in Eq.\eqref{H} and restrict the analysis to translationally-invariant sector $k=0$ with positive parity reflection symmetry. 
As a technical tool, we use the smoothed average of our energy-resolved data as
\begin{equation}
\label{eq_Smooth}
	\overline{[f(x)]}_{\tau} = \frac {\sum_n f(x_n)\delta_\tau(x-x_n)}{\sum_n \delta_\tau(x-x_n)}\ ,
\end{equation}
where, $\delta_\tau(x)$ is a smoothed delta functions such that  ${\lim _{\tau \to \infty}\delta_\tau(x) =\delta(x)}$. In the simulations, we chose a Gaussian smoothing 
$\delta_\tau(x)=e^{- \frac {\tau^2}2 x^2}/\sqrt{2\pi /\tau^2}$.

First, we establish that the initial states \eqref{eq_psi0} are consistent with the ansatz in Eq.\eqref{eth_state}. In Fig.~\ref{Fig1}a we plot the 
diagonal ensemble for different length sizes $L=12, 14, 16, 18$, showing that it
obeys the large deviation prediction {$\overline{\Psi_{ii}} =\frac{e^{-L \phi(E_i/L)}}{Z} $.}
This is confirmed by the inset, where we plot the scaling of the initial energy $E_0=e_0 L$ and variance $\Delta_{E_0}/E_0= \delta_{e_0} / e_0 \sqrt L$, c.f.~Eq.\eqref{Vari}. From a fit of the data, we extract the dimensionless values $e_0 = 0.10$, $\delta_{e_0} = 1.12$.
In panel (b), we study the fluctuations of the out-of-equilibrium ETH functions in the frequency domain [cf. Eq.\eqref{BB}].  To address the dependence on $L$, 
we re-scale $\overline{|\Psi_{ij}|^2}$ by the diagonal matrix elements at energy $e_0$ defined in Eq.\eqref{Scaling_energy}. For the state under consideration, the ansatz in Eq.\eqref{BB} is given by Eq.\eqref{eq_BB_variance} and it is given by
\begin{equation}
	\label{BB_action}
 \frac{\overline{\Psi_{ij}\Psi_{ji}}}{\overline{{|\Psi_{e_0 e_0}|}}^2}  \simeq  e^{- \frac {\phi''(e_0)}{4 L}\omega_{ij}^2} \ .
\end{equation}
In Fig.\ref{Fig1}b, we fix the energy density to be $e_0$ by rescricting the energy indices $i,j$ of $\overline{\Psi_{ij}\Psi_{ji}}$ to $|(E_i+E_j)/2-E_0|\leq \sqrt L \delta_{e_0}$.
The figure shows the smoothed average \eqref{BB_action} as a function of the energy difference $\omega_{ij}=E_i-E_j$ rescaled by $\sqrt L$, for different system sizes. For $L=16$ we also show individual values without smoothing (blue dots). The plot confirms that this initial state has fluctuations that decay as a Gaussian with a variance $1/\sqrt L$, consistent with Eq.\eqref{BB_action}. In the inset, we also confirm the Gaussian decay of the fidelity upon increasing system size [cf. Eq.\eqref{eq_gaussianFidelity}]:  we plot $e^{-\delta_{e_0}^2 t^2}$ with $\delta_{e_0}=1.12$ without fitting parameter.\\

\begin{figure}[t]
\centering
\includegraphics[width=.8\linewidth]{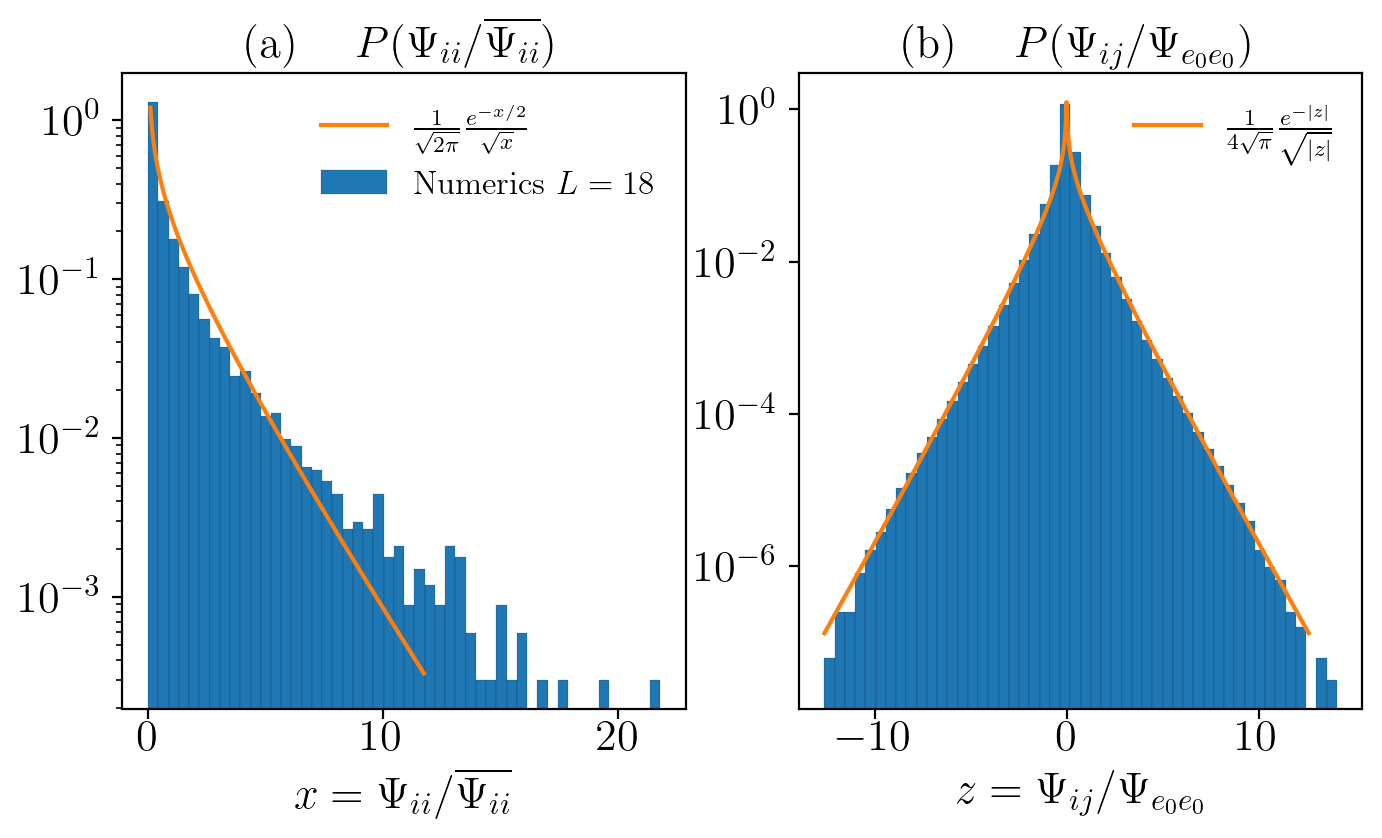}
\caption{ Numerical distributions of the projector's matrix elements $P(\tilde R_{ij})$ for $L=18$ superimposed with the theoretical predictions. (a) Diagonal matrix elements $x=\Psi_{ii}/\overline{\Psi_{ii}}$ (blue) and  the Porter-Thomas distribution in Eq.\eqref{pt} (orange). (b) Off-diagonal matrix elements $z=\Psi_{ij}/\overline{\Psi_{e_0 e_0}}$ (blue) superimposed with the modified Bessel function Eq.~\eqref{eq_bessel} (orange). The data is for the mean energy density $e_0=0.10$ smoothed with  $\tau=4$. 
 }
	\label{fig_2a}
\end{figure}
We show the distribution of  matrix elements $\tilde R_{ij}$ in Fig.~\ref{fig_2a}. In panel (a), we show rescaled diagonal matrix elements $x=\Psi_{ii}/\overline{\Psi_{ii}}$,  $\Psi_{ii}=|c_i|^2$. The data is in agreement with the Porter-Thomas distribution \cite{porter1056fluctuation}, which is  expected for systems with time-reversal symmetry,  falling in the orthogonal ensemble (OE) universality class, 
\begin{equation}
    \label{pt}
    P_{\rm OE}^{}(x) = \frac{e^{-x/2}}{\sqrt{2\pi x}}\ .
\end{equation}
In panel (b), we report the data for the off-diagonal matrix elements, rescaled as  $z=\Psi_{ij}/\overline{\Psi_{e_0 e_0}}$, where $\Psi_{e_0 e_0}$ is defined in Eq.~\eqref{BB_action}. Our numerical results are in  good agreement with Eq.~\eqref{eq_bessel}, without any fitting parameter.

 \begin{figure}[h]
\centering
\includegraphics[width=1\linewidth]{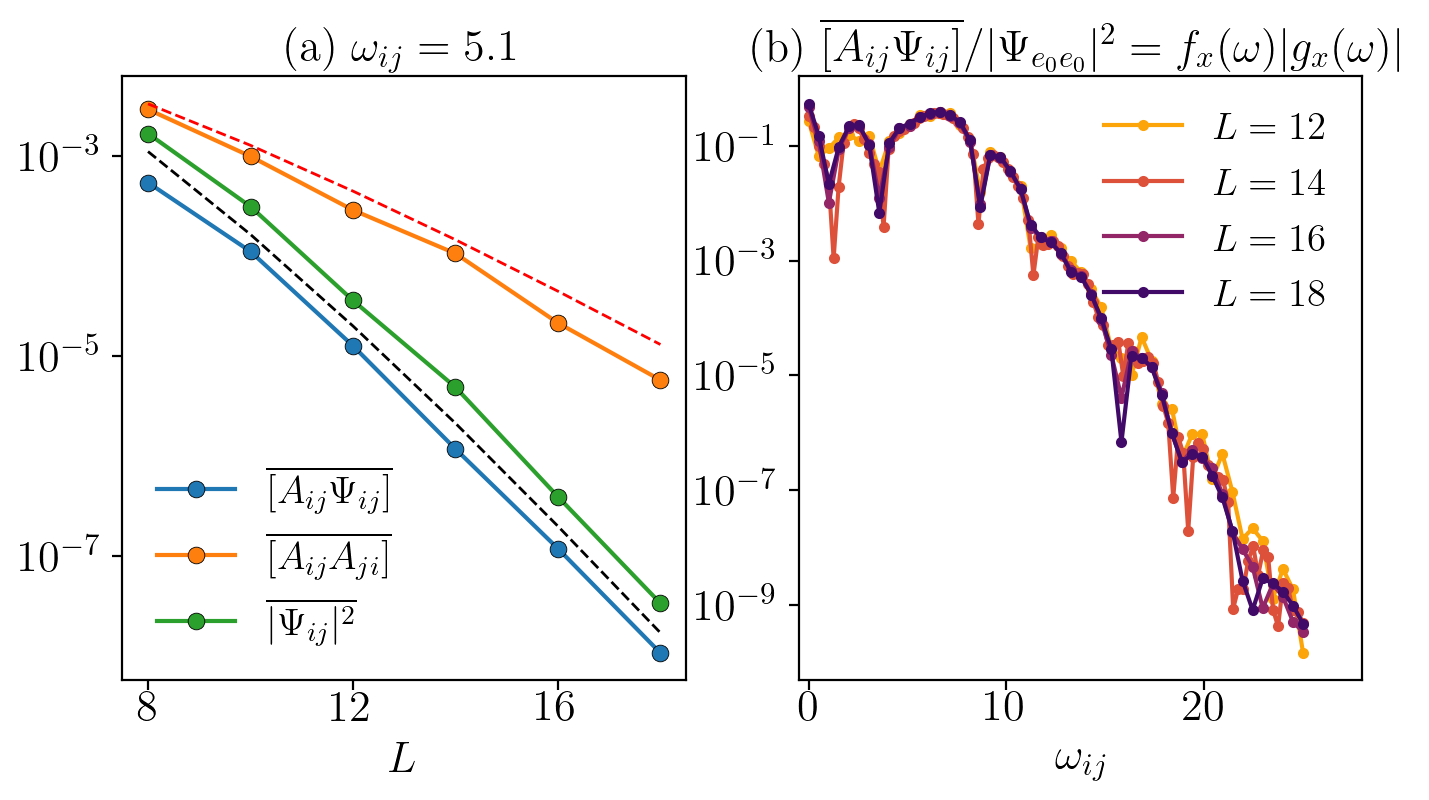}
\caption{ETH correlations between the initial state and the observable $A=\hat \sigma^x_1$. (a) Scaling with the system size of the ETH predictions in Eqs.\eqref{outofeqETH}. The red and black dashed lines indicate $(\dim \mathcal H)^{-1}$ and $(\dim \mathcal H)^{-2}$ respectively  (b)  Smoothed averages $\overline{A_{ij}\Psi_{ij}}$ describing the Fourier transform of the relaxation dynamics increasing system size. The data is for the mean energy density $e_0=0.10$ smoothed with  $\tau=4$. 
 }
	\label{fig_2}
\end{figure}

We then proceed to establish the validity of the ansatz for the correlations between the initial state and observable $\hat A$ in the energy eigenbasis.
In Fig.\ref{fig_2}, we focus on ${|\psi\rangle=|\downarrow_z\dots \downarrow_z\rangle}$ and $\hat A=\hat \sigma_{1}^x$. In panel (a), we test the system size dependency of Eqs.\eqref{outofeqETH} at energy density $e_0$ for finite frequency $\omega_{ij}=5.1$. As predicted by out-of-equilibrium ETH Eq.\eqref{AA}, the observable off-diagonal matrix elements $\overline{A_{ij}A_{ji}}$ decay as $O(e^{-L s(e_0)})$, while both $\overline{|\Psi_{ij}|^2}$ and $\overline{A_{ij}\Psi_{ji}}$ decay as $O(e^{-2 L})$,  cf. Eqs\eqref{BB}-\eqref{AB}. The red and black dashed lines indicate $(\dim \mathcal H)^{-1}$ and $(\dim \mathcal H)^{-2}$ respectively. We checked that the same results hold at zero or for other finite $\omega_{ij}$. \\
In Fig.\ref{fig_2}b, we consider 
\begin{equation}
	\label{fg}
 \frac{\overline{A_{ij}\Psi_{ij}}}{ \overline{\Psi_{e_0e_0}}^2}  \simeq f_A(e_0, \omega)  g_{A, \Psi}(e_0, \omega_{ij})\ ,
\end{equation}
where $\Psi_{e_0e_0}$ is the same as in Eq.\eqref{BB_action} and the right-end side follows from Eq.\eqref{AB} and \eqref{largeDV} \footnote{Using that the diagonal ensemble \eqref{largeDV} for $e^+=e_0$ reads ${|c_{e_0}|^2=\frac {e^{- \Phi(E_0)}}Z \simeq e^{- S(E_0)}}$ where one uses the normalization in Eq.\eqref{norma}}. 
This quantity is of order one, i.e.~it remains finite in the thermodynamic limit.  Its Fourier transform yields equilibration dynamics, see Eq.\eqref{dAt}. 
Note that we have plotted the absolute value of Eq.\eqref{fg}, since the sign of $g(e, \omega)$ oscillates  and this gives rise to the spikes in the curve. This sign change is a characteristic feature of the out-of-equilibrium dynamics, as was emphasized above, as the Fourier transform of $\langle \hat A(t)\rangle$ need not to be necessarily positive. Further investigation will be useful to determine the physical content of these oscillations. 

\begin{figure}[htb!]
\centering
\includegraphics[width=1\linewidth]{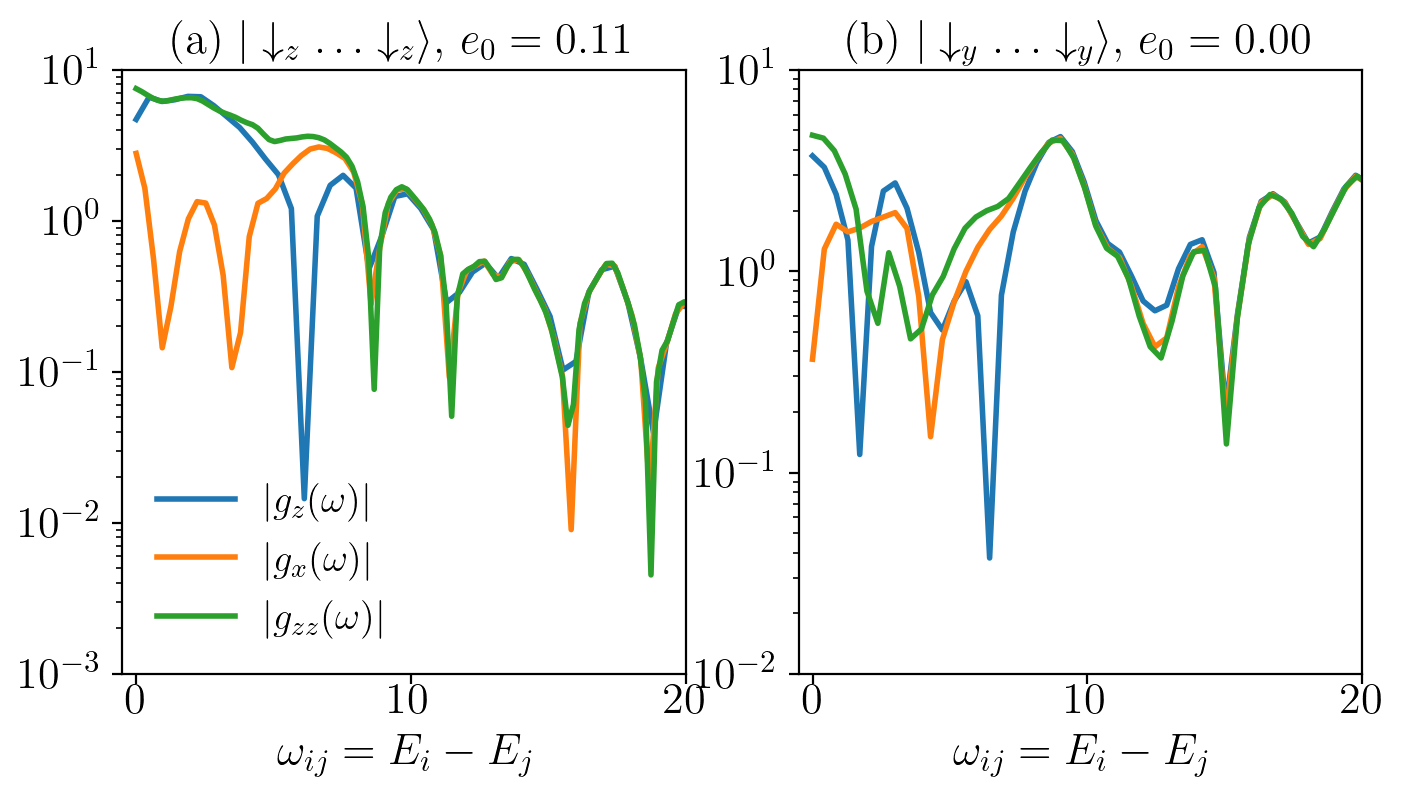}
\caption{Absolute value of the correlations between initial state and observable $|g_{A, \Psi}(e_0, \omega)|$ extracted using Eq.\eqref{eq_g} for different observables $A=\sigma_1^x$, $\sigma_1^z$ and $\sigma_1^{z}\sigma_2^{z}$ as a function of frequency. (a) Results from the initial state $\ket{\downarrow_z \dots \downarrow_z}$. (b) Results from the initial state $\ket{\downarrow_y \dots \downarrow_y}$. Here $L=18$ and $\tau=4$.
 }
\label{fig_3}
\end{figure}

To better understand the behaviour of the function $g(e_0, \omega)$, in Fig.\ref{fig_3} we consider:
\begin{equation}
  	\label{eq_g}
 \frac{\overline{A_{ij}\Psi_{ij}}}{\sqrt{\overline{A_{ij}^2} } \, \sqrt{\overline{|\Psi_{ij}|^2}} 
 \sqrt{\overline{\Psi_{e_0e_0}}}
 }  \simeq  g_{A, \Psi}(e_0, \omega_{ij})\ ,
\end{equation}
to obtain an order one quantity, which encodes the correlations in Eq.\eqref{defg}.
The results are shown in Fig.\ref{fig_3} for the fully polarized states in Eq.\eqref{eq_psi0} along the directions $\alpha=z, y$ in panels (a) and (b) 
respectively, or the three different operators in Eq.\eqref{obs}.  The plot shows that the $g_{A, \Psi}(e_0, \omega)$ may still decay, albeit slowly, as a function of frequency. The most notable fact is that, for different observables, the smooth functions $g$ have approximately the same behavior at large frequencies, which does not depend on the observable. This seems to indicate that the large frequency behavior and the oscillations in the $g_{\psi}(e_0, \omega)$ reflect physics \emph{of the initial state}. \\

\section{Discussion and Conclusions}

In this paper, we have introduced a new ansatz for out-of-equilibrium dynamics, which predicts correlations between the initial state and observables when written in the energy eigenbasis. 

Let us remark that our results describe \emph{a wide class of initial states}, including for example products $|a\rangle_A |b\rangle_B$ of energy eigenstates $|a\rangle_A$ and $|b\rangle_B$ of subsystems $A$ and $B$, that have recently motivated studies of the eigenstate correlations \cite{huang2019universal,PhysRevE.99.032111,murthy2019structure,shi2023local,hahn2023statistical, jindal2024generalized}.
For local $1d$ Hamiltonians, these states have energy fluctuations $\Delta_E^2=O(1)$ in $N$. While obeying Eq.\eqref{Scaling_energy}, they do not have the form of a large deviation \eqref{ext} and their survival probability \eqref{eq_fidelity} is known to decay exponentially in time \cite{torres2014local}. Nevertheless, this class of states is naturally included in our ansatz on the state-observable correlations in Eqs.\eqref{outofeqETH}, where Eq.\eqref{BB} generically reads \begin{equation}
    \overline{|\Psi_{ij}|^2} = e^{- \Phi(E^++\omega/2)-\Phi(E^+-\omega/2)}/Z^2.
\end{equation} These states are discussed in the Appendix, where we verify numerically the general scaling with the system size of Eqs.\eqref{outofeqETH}, and comment on the relation with the literature. \\

Our work opens a series of perspectives.
At long times, hydrodynamic modes are expected to play a dominant role in equilibration dynamics \cite{mukerjee2006statistical,lux2014hydrodynamic,michailidis2023corrections}, and it would be valuable to investigate how hydrodynamic description can be incorporated into the non-equilibrium ETH. Additionally, one could explore how the current ETH framework applies to integrable systems that equilibrate to a generalized Gibbs ensemble \cite{alba2015eigenstate, essler2016quench, vidmar2016generalized, essler2023statistics} or in the presence of many-body quantum scars \cite{
serbyn2021quantum,moudgalya2022quantum,chandran2023qmbsreview, gotta2023asymptotic}. 

\vspace{1cm}

\section*{Acknowledgements}

We thank J. Kurchan for inspiring discussions and collaboration on related topics.   We thank A. Rosch and D. Abanin for useful discussions. 
AD  is supported by the NSF grant PHY 2310426.  S.P. acknowledges support by the Deutsche Forschungsgemeinschaft (DFG, German Research Foundation) under Germany’s Excellence Strategy - Cluster of Excellence Matter and Light for Quantum Computing (ML4Q) EXC 2004/1 -390534769.

\begin{appendix}

\section{Additional statistical insights}
\subsection{A simple scaling drawn from RMT}
\label{app_sca}

Let us discuss a simple example that exhibits the scaling in Eq.\eqref{outofeqETH}. Consider an observable $A= \sum_{\alpha} \lambda_{\alpha} |\lambda_{\alpha}\rangle \langle \lambda_{\alpha}|$ and as an initial state we will take an eigenvector of such observable $|\psi\rangle = |\lambda_{\psi}\rangle$.
Let us suppose that afterward, the state evolves under a  ${\mathcal D}\times {\mathcal D}$ Hamiltonian that is drawn from a rotationally invariant ensemble, i.e. $P(H) = P(U^{-1} H U)$ where $U$ is arbitrary  orthogonal (or unitary) matrix, for instance a GOE or GUE ensemble. With this choice, the Hamiltonian eigenvectors $|E_j\rangle$, in the basis of the observable, i.e. $\langle E_j |\lambda_\alpha\rangle$, are represented by random orthogonal or unitary matrices. 
The properties of the matrix elements of a given observable $A$ in such random basis have been discussed in \cite{High_T}.
In the large ${\mathcal D}$ limit assuming to initialise the dynamics in $|\psi \rangle  = |\lambda_{\psi}\rangle$, some eigenvector of the observable $A$,
the properties of this toy ``out-of-equilibrium" ETH can be easily derived 
\begin{subequations}
\label{outofeqETH_RMT}
\begin{align}
	\label{AA_RMT}
	\overline{|A_{ij}|^2} & = \frac{1}{{\mathcal D}} \, (\langle A^2\rangle - \langle A\rangle^2)
	\\ 	
	\overline{|\Psi_{ij}|^2 } & = \frac{1}{{\mathcal D}^2} \qquad \text{for $i\neq j$}
	\\
	\overline{\Psi_{ij} A_{ij}} & = \frac{1}{{\mathcal D}^2} \left[ \lambda_{\psi} - \langle A \rangle  \right]
\end{align}
\end{subequations}
where $\langle \bullet\rangle = \frac{1}{{\mathcal D}} \Tr(\bullet)$. This is a particularly simple example of the ansatz  \eqref{outofeqETH} discussed in the main text.  In 
term of normalized fluctuations this means $\overline{R_{ij}\tilde{R}_{ij}} \simeq {\mathcal D}^{-{1/2}}$, as in (\ref{defg}).
This examples illustrates the difference in scaling between
$\overline{|A_{ij}|^2}$ and $\overline{|\Psi_{ij}|^2 }$. The first quantity has rank ${\mathcal D}$, leading to $\sum_{ij} \overline{|A_{ij}|^2} = O({\mathcal D})$, while the second has rank one, $\sum_{ij} \overline{|\Psi_{ij}|^2} = O(1)$. Similarly  $\sum_{ij} \overline{\Psi_{ij} A_{ij}} = O(1)$,  which is consistent with the scaling above.

\subsection{A constraint on the ansatz}\label{App_norm}

Let us justify Eqs. (\ref{Eq_RRR}). As we stressed several times, contrary to the standard ETH ansatz for observables, the matrix that we have are chracterising has rank 1.
In particular, calling $z_i = \frac{1}{\sqrt{Z}} e^{-\frac{1}{2} \Phi(E_i)}$ and following the notation of the main text we have:
\begin{equation}
\begin{array}{c}
\displaystyle
1 + \tilde{R}_{ii} = \frac{|c_i|^2}{z_i^2}  
\\ \vspace{-0.2cm} \\
\displaystyle
\tilde{R}_{ij} = \frac{c_i c_j^{\ast} }{ z_i z_j }  \qquad \text{for $i \neq j$}
\end{array}
\end{equation}
Taking products:
\begin{equation}
\label{ikj}
\tilde{R}_{ik} \tilde{R}_{kj} = \frac{c_i |c_k|^2 c_j^{\ast}}{z_i |z_k|^2 z_j} =  \tilde{R}_{ij} (1+\tilde{R}_{kk})  \qquad \text{for $i \neq j \neq k$}
\end{equation}
and similarly
\begin{equation}
\label{ikik}
\tilde{R}_{ij} \tilde{R}_{ji} = \frac{|c_i|^2 |c_j|^2}{|z_i|^2 |z_j|^2} = (1 + \tilde{R}_{ii})(1+\tilde{R}_{jj})  \qquad \text{for $i \neq j$}
\end{equation}

Let us now see how these constraints imply that $\Psi$ is a projector by proving Eqs. (\ref{Eq_Psi2}).
We start by evaluating the off-diagonal with $i\neq j$:
\begin{align}
\begin{split}
    [\Psi^2]_{ij} & = \sum_{k} \Psi_{ik}\Psi_{kj} = \Psi_{ii}\Psi_{ij} + \Psi_{ij }\Psi_{jj} + \sum_{k: k\neq i\neq j} \Psi_{ik} \Psi_{kj}\\
    & =
    \frac 1{Z^2} e^{- \frac 12 (\Phi(E_i) + \Phi(E_j) )} \left [ e^{-\Phi(E_i)} \tilde R_{ii} \tilde R_{ij} +  e^{-\Phi(E_j)} \tilde R_{jj} \tilde R_{ij}\right ]
    +\frac 1{Z} e^{- \frac 12 (\Phi(E_i) + \Phi(E_j) )} \sum_{k: k\neq i \neq j} \frac{e^{- \Phi(E_k) }}Z \tilde R_{ik} \tilde R_{kj}
    \end{split}
\end{align}
where from the first to the second line we have inserted the ansatz \eqref{outofeqETH} in the individual matrix elements. The first term is subleading $O(e^{-2N})$, while in the second term, we can substitute the ansatz of Eq.\eqref{ikj} and obtain
\begin{align}
\begin{split}
    [\Psi^2]_{ij} & \simeq \frac 1{Z} e^{- \frac 12 (\Phi(E_i) + \Phi(E_j) )} \sum_{k: k\neq i \neq j} \frac{e^{- \Phi(E_k) }}Z \tilde R_{ij} (1+\tilde R_{kk})
     =  \frac 1{Z} e^{- \frac 12 (\Phi(E_i) + \Phi(E_j) )}  \tilde R_{ij}\sum_{k: k\neq i \neq j} \Psi_{kk} 
     \\ &=  \frac 1{Z} e^{- \frac 12 (\Phi(E_i) + \Phi(E_j) )}  \tilde R_{ij} = \Psi_{ij}\ ,    
\end{split}
\end{align}
where we used $\sum_{k: k\neq i \neq j} \Psi_{kk} \simeq \sum_k \Psi_{kk}=\langle \psi|\psi\rangle = 1$ which shows the first equation in (\ref{Eq_Psi2}). Similar manipulations can be done on the diagonal elements, leading to
\begin{align}
\begin{split}
    [\Psi^2]_{ii} & = \sum_{k} \Psi_{ik}\Psi_{ki} = \Psi_{ii}\Psi_{ii} + \sum_{k: k\neq i} \Psi_{ik} \Psi_{ki}\\
    & =
    \frac {e^{- 2\Phi(E_i) }}{Z^2}  (1+\tilde R_{ii})^2
    +\frac 1{Z} e^{- \Phi(E_i) } \sum_{k: k\neq i} \frac{e^{- \Phi(E_k) }}Z \tilde R_{ik} \tilde R_{ki}
    \\ & \simeq \frac 1{Z} e^{- \Phi(E_i) } (1+\tilde R_{ii})  \sum_{k: k\neq i} \frac{e^{- \Phi(E_k) }}Z (1+\tilde R_{kk}) = \Psi_{ii} \sum_{k\neq i} \Psi_{kk}
    = \Psi_{ii}\ ,
    \end{split}
\end{align}
where, from the second to the third line we have used the fact that the first term is subleading and the ansatz in Eq.\eqref{ikik}.

In addition to \eqref{ikj} and \eqref{ikik}, there is an infinite number of higher-order constrains coming from the fact that $\Psi_{ij}$ is a projector. 
Say, we find at order $3$,
\begin{equation}
\tilde{R}_{ij}\tilde{R}_{jk} \tilde{R}_{ki}  = \tilde{R}_{ik} (1+ \tilde{R}_{jj} )\tilde{R}_{ki} = (1+\tilde{R}_{ii})(1+\tilde{R}_{jj})(1+\tilde{R}_{kk}),
\end{equation}
for $i\neq j\neq k$
as well as 
 for $i\neq j\neq k\neq l$ at order $4$,
\begin{equation}
\tilde{R}_{ij}\tilde{R}_{jk} \tilde{R}_{kl}\tilde{R}_{li}  = \tilde{R}_{ik} (1+ \tilde{R}_{jj} )\tilde{R}_{ki}(1+\tilde{R}_{ll} ) = (1+\tilde{R}_{ii})(1+\tilde{R}_{jj})(1+\tilde{R}_{kk})(1+\tilde{R}_{ll}).
\end{equation}
The list of such constraints continues to include higher orders of $\tilde{R}$.

\subsection{Fluctuations of state-observable correlations }

At the level of single matrix elements, the product of the initial state and the observable has large fluctuations in the system size. In fact
\begin{equation}
    \Psi_{ij} A_{ij} \simeq \overline{\Psi_{ij}A_{ji}}  + \sqrt{\overline{|\Psi_{ij}|^2}} \, \sqrt{\overline{|A_{ij}|^2}} \, \xi_{ij}
\end{equation}
with $\xi_{ij}$ some random variable with average zero and fluctuations order one. Here, the amplitude of the fluctuations is larger than the average:
$$\sqrt{\overline{|\Psi_{ij}|^2} \, \overline{|A_{ij}|^2}} \gg \overline{\Psi_{ij}A_{ji}}\ ,$$
since $\sqrt{\overline{|\Psi_{ij}|^2} \, \overline{|A_{ij}|^2}}\simeq e^{-3S/2}$ and $\overline{\Psi_{ij}A_{ji}}\sim e^{-2 S}$.
However, when computing physical observables, one has to sum over many indices, and, due to the presence of randomness, the fluctuations become negligible.
This is analogous to what happens to high-order products of matrix elements in standard ETH, which also possess large fluctuations \cite{FK2019}. 
These fluctuations contribute, at least for finite systems sizes, to $\langle A(t) \rangle \langle A(-t) \rangle $, 
A detailed understanding of their influence on the dynamics is left to future work.

\section{A bound on high frequency tail of $g(\omega)$}\label{ghft}

As a starting point, we introduce
\begin{equation}
\label{Zb}
Z(\beta)=\sum_i e^{-\Phi(E_i)-\beta E_i}
\end{equation}
such that  $Z=Z(0)$ matches the definition of $Z$ in \eqref{eth_state}. It is also convenient to introduce the ratio
\begin{equation}
    {\cal Z}(\beta)=\langle \psi|e^{-\beta H}|\psi\rangle={Z(\beta)\over Z}.
\end{equation}

To constrain $g_{\psi,A}$ we use the approach similar to  one used \cite{avdoshkin2020euclidean}, which 
bounds on high-frequency tail of $f_A$, 
\begin{equation}\label{bound_f}
    |f_A(\tilde{e},\omega)|\leq O\left(e^{-(\tilde{\beta}/4+\beta^*)\omega}\right),\qquad
    \omega\rightarrow \infty.
\end{equation}
where $\beta^*$ is an $O(1)$ constant defined by local model parameters. 
Here $O\left(\dots \right)$ means that possible pre-exponential $\omega$-dependent factors are ignored.
Finally,    temperature $\tilde{\beta}$ is associated with energy  density $\tilde{e}$, $S'(N\tilde{e})=\tilde{\beta}$.
We now consider 
\begin{equation}
\label{Cpsi}
  C_\Psi^\beta(t)\equiv {  \langle \psi|e^{-\beta H}A(t)|\psi\rangle\over   {{\cal Z}}(\beta)}.
\end{equation}
We now use the following inequality  
\begin{equation}
|\langle \psi|A|\psi'\rangle|\leq |A| |\psi||\psi'|,
\end{equation}
where $|A|$ is an infinity norm of the operator $A$, meaning the largest (by absolute value) eigenvalue when $A$ is hermitian, or largest singular value when $A$ is not hermitian. Taking $|\psi\rangle= \ket \psi$ and $\bra{\psi'}=\bra{\psi}e^{-\beta H}$ we arrive at 
\begin{equation}
 \label{bb}
     \left|\int d\omega\,  f_A(\tilde{e},\omega)g_{\psi,A}(\tilde{e},\omega) e^{\omega (it-\beta/2)}\right| \leq |A(t)| {Z^{1/2}(2\beta) Z^{1/2}(0)\over Z(\beta)}.
\end{equation}
Here $\tilde{e}$ is the energy density where the main contribution to the integral in \eqref{Zb} comes from, $\tilde{e}=-\partial/\partial \beta\ln Z(\beta)/N$.
We can now redefine $t \rightarrow t-i\beta/2$,
\begin{equation}
 \label{bb1}
     \left|\int d\omega\,  f_A(\tilde{e},\omega)g_{\psi,A}(\tilde{e},\omega) e^{i\omega t}\right| \leq |A(t-i\beta/2)| {Z^{1/2}(2\beta) Z^{1/2}(0)\over Z(\beta)}.
\end{equation}
The LHS is an even function of $t$, while $|A(t)|$ is analytic within the strip $|\Im(t)|\leq \beta^*$ \cite{avdoshkin2020euclidean}. Thus the RHS of \eqref{bb1} is analytic inside the strip $-\beta^*+\beta/2\leq\Im(t)\leq \beta^*+\beta/2$. Because the LHS is even, it has to be analytic inside a wider strip $-\beta^*-\beta/2\leq\Im(t)\leq \beta^*+\beta/2$. For the integral over $\omega$ to converge, taking into account the bound \eqref{bound_f} we find
\begin{equation}
\label{bound}
    |g_{\psi,A}(\tilde{e},\omega)|\leq O\left(e^{-(\beta/2 -\tilde{\beta}/4)\omega}\right){Z^{1/2}(2\beta)Z^{1/2}(0)\over Z(\beta)},\qquad
    \omega\rightarrow \infty.
\end{equation}
Here $\beta$ is a free parameter, it determines the ``saddle point'' (mean energy density) $\tilde{e}(\beta)$, where the integral in  \eqref{Zb} is saturated, which in turn defines $\tilde{\beta}$. Parameter $\beta^*$, which characterizes the model does not appear in \eqref{bound}. 
When $\beta=0$, mean energy density $\tilde{e}=e_0$, and $\tilde{\beta}$ is the effective temperature of state $\psi$. 

The logic behind free parameter 
 $\beta$ appearing in bound  \eqref{bound} is exactly as in \cite{avdoshkin2020euclidean}, this is a parameter to optimize over, to find the best possible bound. For the large deviation states \eqref{ext}, the factor ${Z^{1/2}(2\beta)\over Z(\beta)}$ grows extensively, $\ln({Z^{1/2}(2\beta)\over Z(\beta)})\sim N\, O(\beta^2)$, not leading to a meaningful bound in the thermodynamic limit. 

For the states with energy of order one, e.g.~discussed in the next section, effective energy density $\tilde{e}=e_0$ is $\beta$-independent, $\tilde{\beta}=\beta_0$, at least so for  the parameter $\beta$ smaller than certain value $\beta<\lambda$, see \eqref{eq_Fpred}.
In this regime ${Z^{1/2}(2\beta)Z^{1/2}(0)/Z(\beta)}$ is of order one, and $g_{\psi,A}(\tilde{e},\omega)$ for large $\omega$ decays exponentially, bounded by $e^{-(\lambda/2-\beta_0/4)\omega}$, provided $\beta_0\leq 2\lambda$.

\section{Bipartite energy eigenstates}
Consider a tensor product Hilbert space $\mathcal H = \mathcal H_A\otimes \mathcal H_B$ and a Hamiltonian 
$ \hat H = \hat H_A + \hat H_B + \hat H_{AB}$, describing an interaction of a (sub)system $A$ with a ``bath'' $B$. A particular example to keep in mind is a spin-chain split into two parts,  interacting through 
a local term 
$H_{AB}$. We start with a pair 
$\ket{a}$, $\ket{b}$ of the  eigenstates of $\hat H_A$, $\hat H_B$ respectively 
and consider a  state $|\psi\rangle=|ab\rangle$. Decomposition of this state in the eigenbasis  $\ket{E_i}$ of $\hat H$ defines the projector \eqref{eth_state}, 
\begin{equation}
    \label{olapBIPA}
     \Psi^{(ab)}_{ij}= \langle E_i |ab\rangle  \langle ab |E_i\rangle  = c^{(ab)}_i  c^{(ab) *}_j\ .
\end{equation}
These initial states
have been the focus of a great attention lately \cite{Deutsch_2010,Dymarsky:2016ntg,huang2019universal,PhysRevE.99.032111,murthy2019structure,shi2023local,hahn2023statistical, jindal2024generalized}, , especially in relation to bipartite entanglement.
The statistical properties of \eqref{olapBIPA} can be formulated in terms of the so-called \emph{Ergodic Bipartition} (EB) ansatz \cite{jindal2024generalized}, which postulates that on average 
\begin{equation}
    \label{EB} \overline{\Psi_{ii}^{(ab)} }
    = \overline{|c^{(ab)}_i|^2} = e^{-S(E_a+E_b)}F(E_i-E_a-E_b)\ ,
\end{equation}
where $F(E_i-E_a-E_b)$ is a narrowly-peaked function around $E_i\simeq E_a+E_b$. More accurately instead of $E_a+E_b$ in the expression above one should use mean energy $E_0$ of $|ab\rangle$, as we do below. 
More detailed properties of $F(x)$ for 1D systems with local interactions, 
the Lorentzian shape  at small $x$ and exponential suppression at large $x$, 
\begin{align}
    \label{eq_Fpred}
    F(x) \propto 
    \left\{
    \begin{array}{lr}
  (x^2 + \Delta^2)^{-1}, & x \ll  \sigma,
    \\
    e^{- |x| \lambda}, & x\gg  \sigma,
    \end{array}\right.
\end{align}
where $\sigma, \Delta, \lambda$ are model-dependent local (finite in the thermodynamic limit) parameters, 
were outlined in \cite{shi2023local}.

From the definition \eqref{olapBIPA} it is  clear the ergodic bipartition ansatz \eqref{EB} is  the diagonal part of the out-of-equilibrium ETH  ansatz
\eqref{eth_state} applied to a particular initial states. 
We now discuss how our approach encompasses the properties of \eqref{EB}. Starting from \eqref{largeDV}, using saddle point approximation we find,
\begin{align}
\label{diagonal}
   \overline{\Psi_{ii}} 
   = \frac{e^{-\Phi(E_i)}}Z 
   \approx e^{-S(E_0)} 
   \frac{e^{-\Phi(E_i)+\Phi(E_0)}}{\sqrt{2\pi}\Delta_{E_0}},
\end{align}
where $E_0$ is the mean energy of state $|ab\rangle$, $E_0=E_a+E_b+\Delta_{ab}$. Here $\Delta_{ab}=\langle ab|\hat{H}_{AB}|ab\rangle$ and $$\Delta_{E_0}^2=\langle ab|\hat{H}_{AB}^2|ab\rangle-(\Delta_{ab})^2 = \mathcal O(1)$$ are of order one, i.e.~remain finite in the thermodynamic limit, while $E_0$ and $E_a+E_b$ are extensive.
From here follows 
\begin{equation}
    F(E_i-E_0) \approx    \frac{e^{-\Phi(E_i)+\Phi(E_0)}}{\sqrt{2\pi} \Delta_{E_0}}
\end{equation}
which is sharply peaked around mean energy $E_i\approx E_0$ with the variance of order one.

Integrating \eqref{diagonal} over $dE_b$ with the weight $e^{S_B(E_b)}$ and the constraint $E_0\approx E_a+E_b$ readily gives diagonal approximation to reduced density matrix \cite{Dymarsky:2016ntg}
\begin{equation}
    {\rm Tr}_B(|E_i\rangle\langle E_i)\approx \int dE_a\, e^{S_A(E_a)} e^{-S(E_i)+S_B(E_i-E_a)} |a\rangle \langle a|,
\end{equation}
from where von Neumann and Renyi entropy follow via saddle point approximation. Justification of the diagonal approximation to evaluate entropy was recently addressed in \cite{jindal2024generalized} by considering the statistical properties of 
\begin{equation}
    c_i^{(ab)}  c^{(a'b) *}_i  c_i^{(a'b')}  c_i^{(ab')*},
\end{equation}
which can be understood extending the ansatz to four different states $\ket{\psi_1}=\ket{ab}$, $\ket{\psi_2}=\ket{a'b}$, $\ket{\psi_3}=\ket{a'b'}$, $\ket{\psi_4}=\ket{ab'}$, with $i_1=i_2=i_3=i_4$.

\begin{figure}[hbt!]
	\centering
	\hspace{-.5cm}
	\includegraphics[width=.45\linewidth]{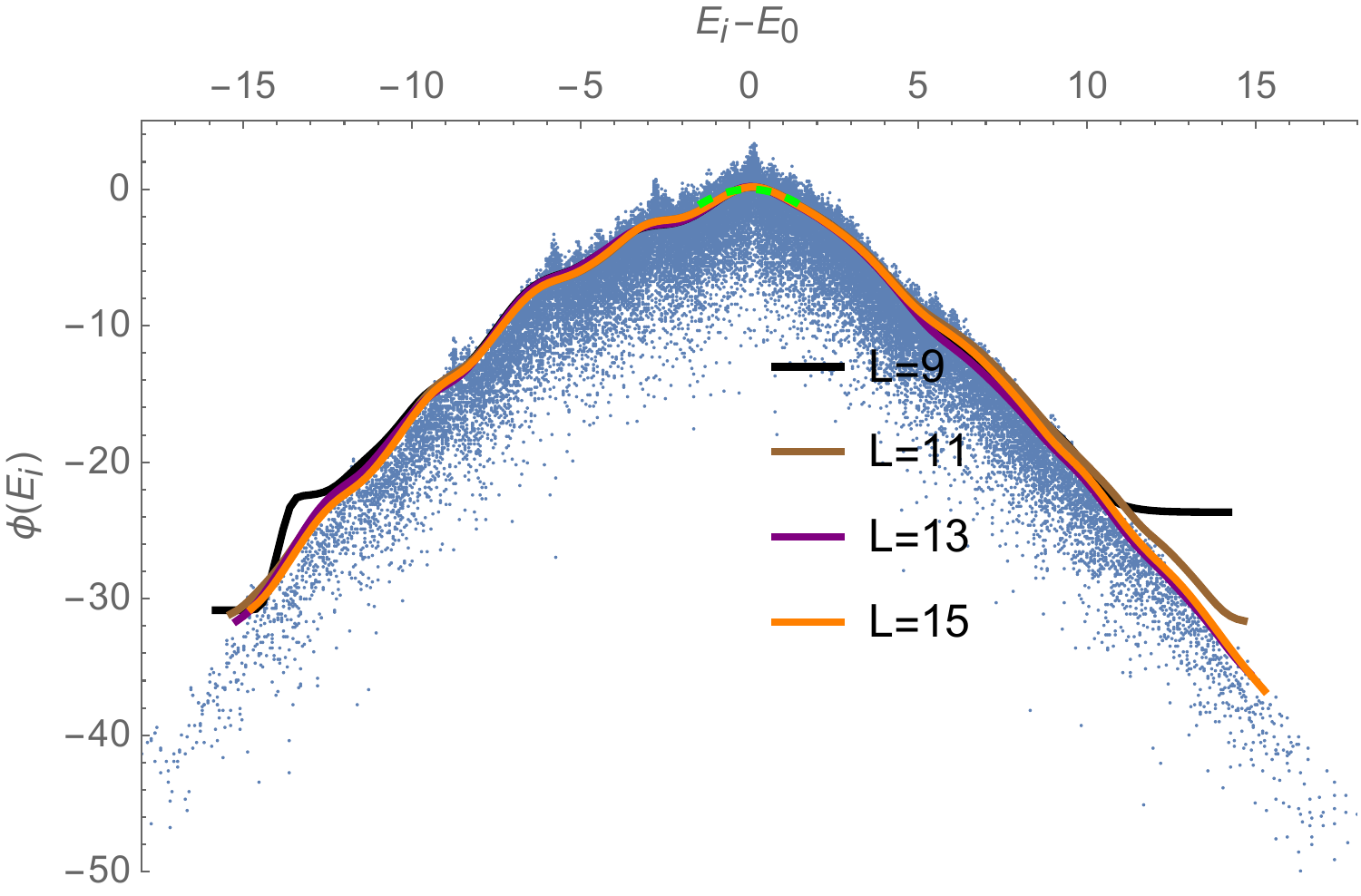}
    \includegraphics[width=.45\linewidth]{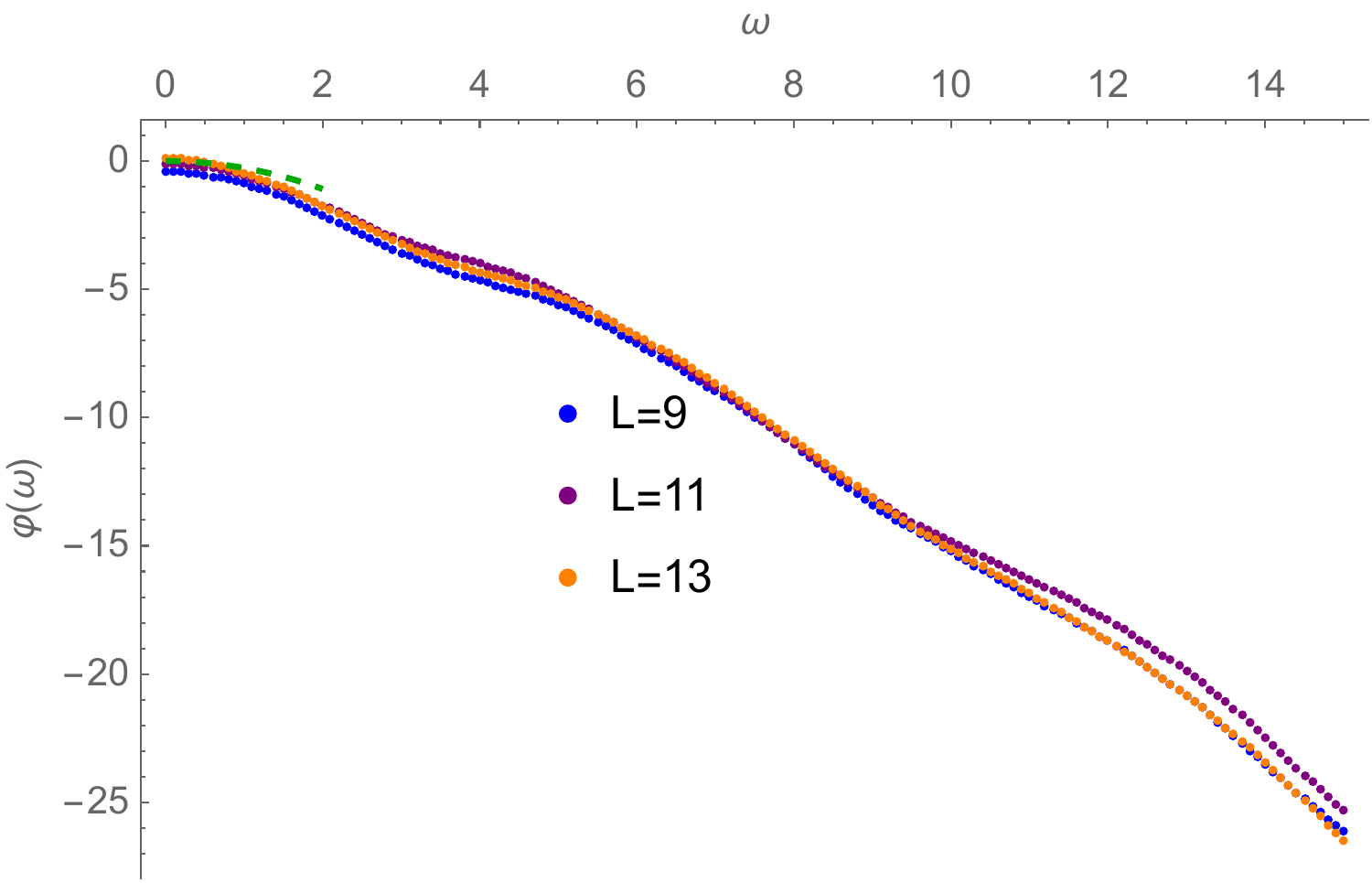}
\caption{Left panel: averaged 
diagonal matrix elements \eqref{diagonal}, with the entropy subtracted $\phi(E_i)\equiv\ln \overline{\Psi_{ii}}+S(E_0)+{1\over 2}\ln (2\pi \Delta E^2)$, for different system sizes, superimposed with ${-(E_i-E_0)^2/( 2\Delta^2_{E_0})}$ with the value of $\Delta^2_{E_0}$ for $L=15$ (green dashed line). Blue points show raw data (un-averaged value of $\ln \Psi_{ii}$, with the entropy subtracted) for $L=15$. 
 Right panel: average of the off-diagonal matrix elements \eqref{S1r} for different system sizes, superimposed with ${-\omega^2/(4\Delta^2_{E_0})}$ with the value of $\Delta^2_{E_0}$ for $L=13$ (green dashed line).
}
	\label{fig_state_Half}
\end{figure}

To illustrate the behavior of $\Phi(E_i)$ we consider, c.f.~\eqref{eq_Fpred}, 
\begin{equation}
\label{Psiexp}
       \ln \overline{\Psi_{ii}} \simeq -S(E_0)- {1\over 2}\ln (2\pi \Delta E^2)-{(E_i-E_0)^2\over 2\Delta^2_{E_0}}+\dots 
\end{equation} 
and note that for $E-E_0$ of order one, higher-order corrections to \eqref{Psiexp} are also of order one and can not be neglected \cite{shi2023local}.
It corresponds to the  first two terms of the Taylor expansion of \eqref{eq_Fpred} in $(E_i-E_0)^2$.
We plot $\phi(E_i)\equiv\ln \overline{\Psi_{ii}}+S(E_0)+{1\over 2}\ln (2\pi \Delta E^2)$ as a function of $E_i-E_0$
numerically in the left panel of Fig.~\ref{fig_state_Half} for the tilted field Ising model \eqref{H} of size $L=2L'+1$ with the parameters $J=-1,w=1.05,h=0.4$. The eigenstates $|a\rangle,|b\rangle$ are chosen to be the ground state and the most excited states of the subsystems of size $L'$ and $L'+1$ correspondingly. With this choice of $\psi_0$ the total energy $E_0$ is very close to the middle of the spectrum.  
To obtain $\overline{\Psi_{ii}}$ we use smoothed average \eqref{eq_Smooth} with $\tau=2$. The entropy 
\begin{eqnarray}
    S(E)=\ln \Omega(E_0), \qquad \Omega(E)={\kappa\, L!\over (L/2-\kappa E)!(L/2+\kappa E)!}
\end{eqnarray}
is evaluated using binomial analytic approximation for the density of states of the titled filed Ising model, see Appendix A of \cite{Dymarsky:2016ntg}, with $\kappa={1\over 2}\sqrt{J^2+w^2+h^2-1/L}$. Mean energy $E_0$ and energy variance $\Delta^2_{E0}$ are evaluated numerically for each $L$. 
The plot in \ref{fig_state_Half} shows good collapse of $\phi(E_i-E_0)$ for different values of $L$, and is well described by ${(E_i-E_0)^2\over 2\Delta^2_{E_0}}$ for small $|E_i-E_0|$.

The out-of-equilibrium ETH anzats predicts the off-diagonal matrix elements 
\begin{equation}
\label{S1r}
 \varphi(\omega)=\ln  \overline{\Psi_{ij}^2} -2\ln  \overline{\Psi_{ee}}
   \simeq  -{\omega^2\over 4\Delta^2_{E_0}},\quad |\omega| \lesssim \Delta_{E_0},
\end{equation}
plotted in the right panel of Fig.~\ref{fig_state_Half}. 
Here $\overline{\Psi_{ee}}$ denotes the average $\overline{\Psi_{ii}}$ over a narrow shell around $E_i=E_0$ of size $\Delta_{E_0}$. Similarly, $\Psi_{ij}^2$ in \eqref{S1r} is averaged only over pairs $E_i,E_j$ satisfying $|(E_i+E_j)/2-E_0|\leq \Delta_{E_0}$.

 \begin{figure}[hbt!]
	\centering
 \hspace{-1cm}
\includegraphics[width=.33\linewidth]{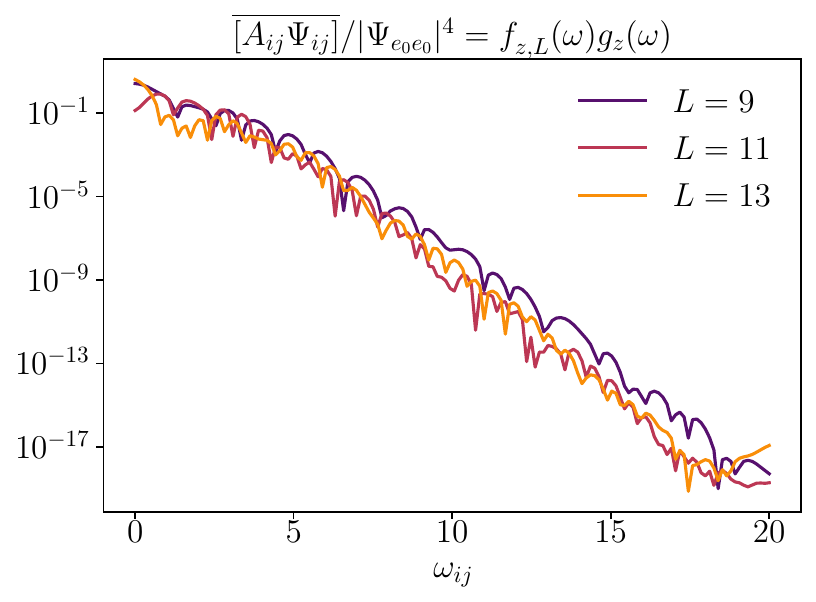}
\includegraphics[width=.33\linewidth]{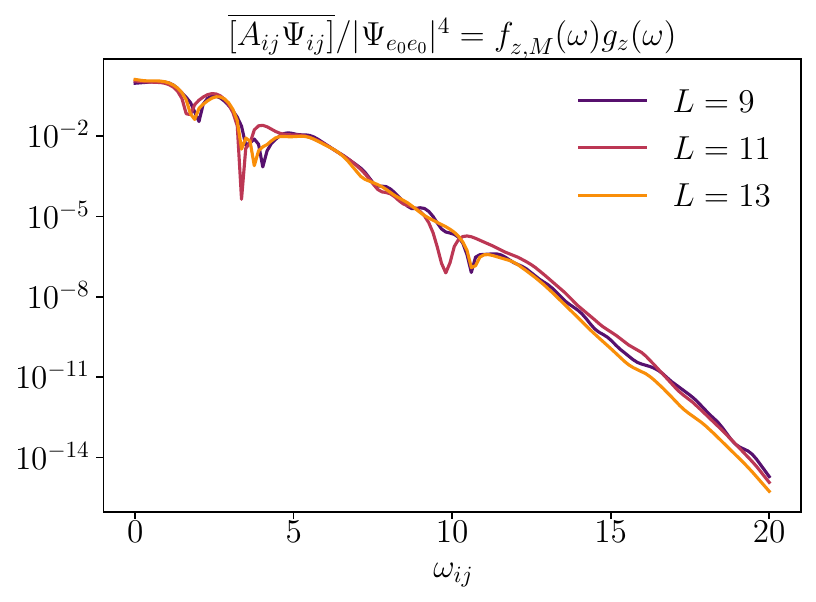}
\includegraphics[width=.33\linewidth]{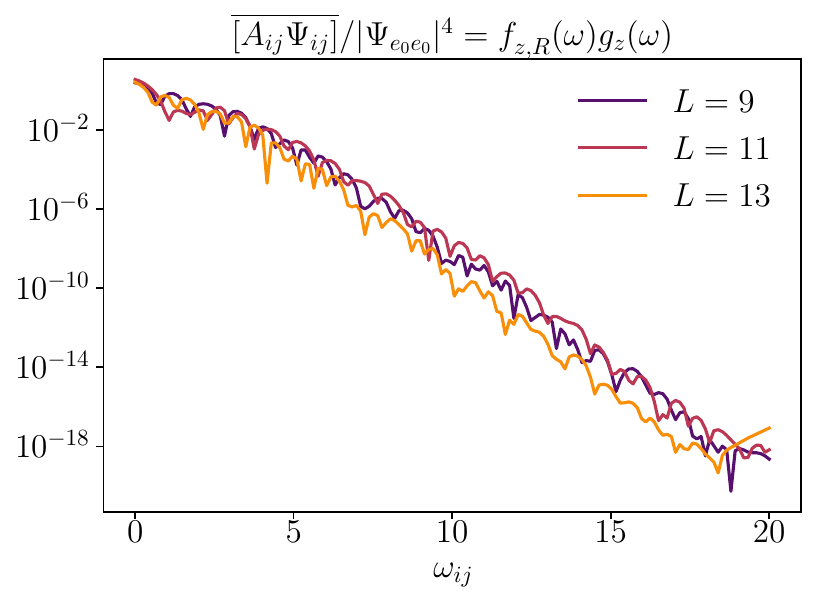}\\
 \hspace{-1cm}
\includegraphics[width=.33\linewidth]{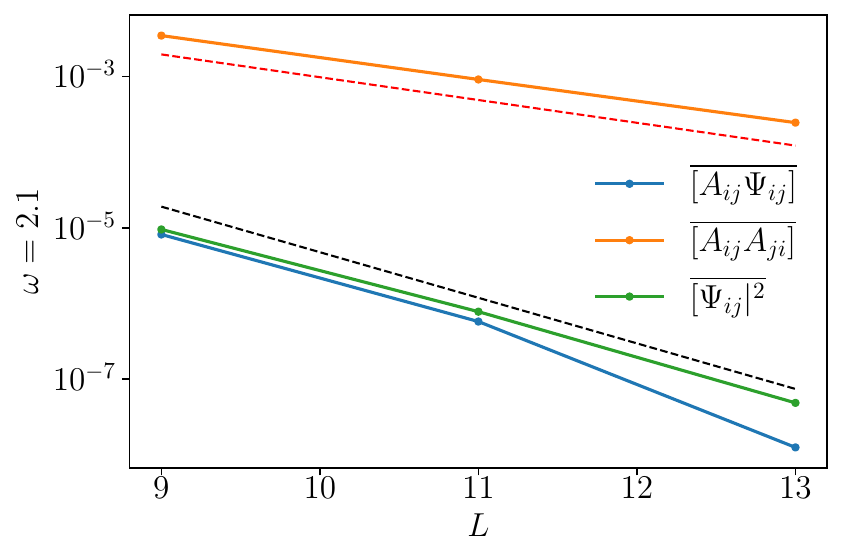}
\includegraphics[width=.33\linewidth]{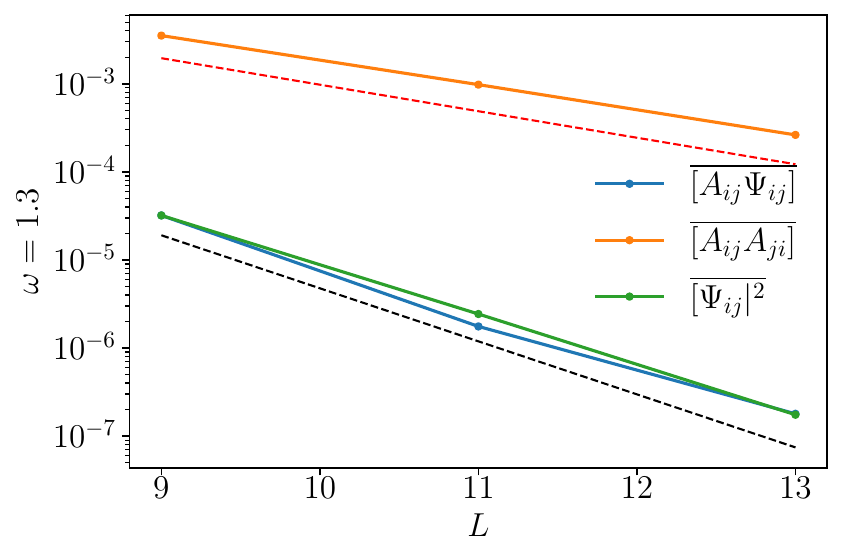}
\includegraphics[width=.33\linewidth]{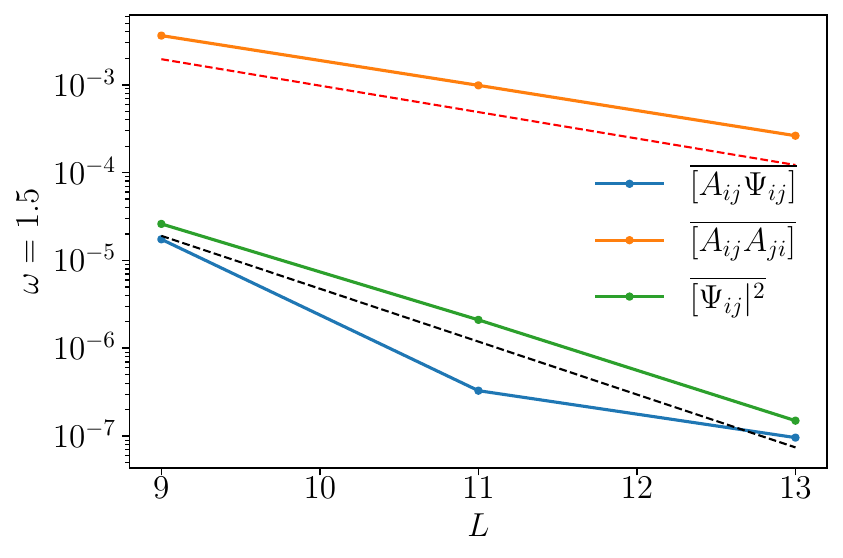}
\caption{First row: from left to right, plots of $f_A(\omega)g_{\psi,A}(\omega)$ for $A=\sigma^z_L,\sigma^z_M,\sigma^z_R$ correspondingly. 
Second row: scaling of Eqs.\eqref{outofeqETH} contrasted with $(\dim \mathcal H)^{-1}$ (dashed red line) and  $(\dim \mathcal H)^{-2}$ (dashed black line).}
	\label{fig_din_zoom_psiHalf}
\end{figure}

\begin{figure}[hbt!]
	\centering
\includegraphics[width=.47\linewidth]{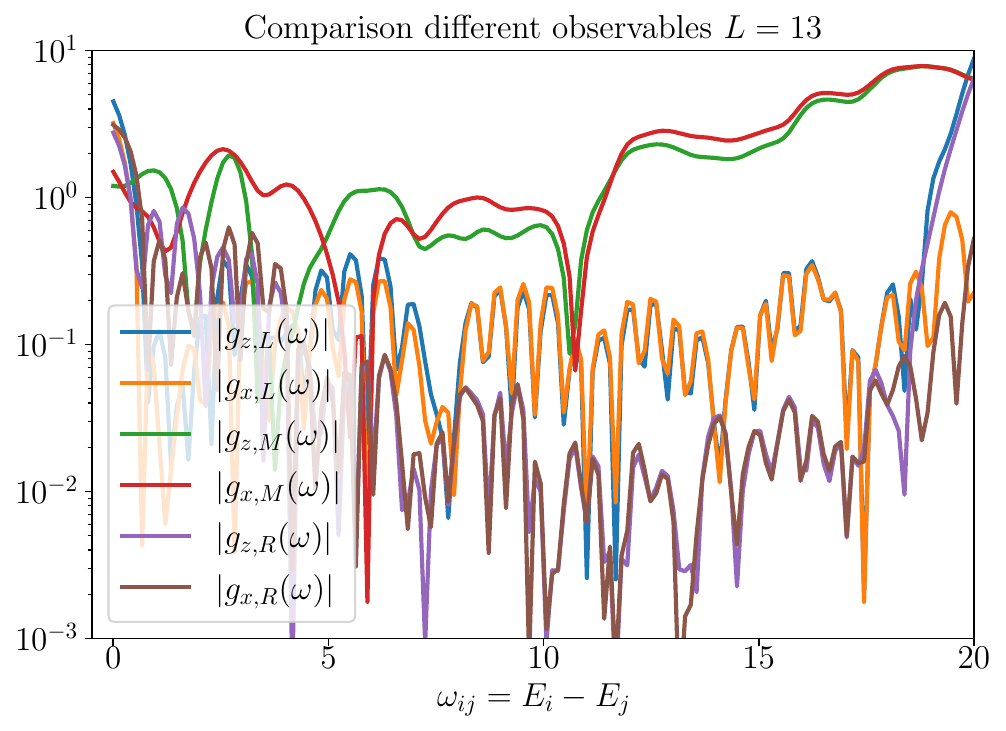}
\caption{In the absence of translational invariance, the function $g$ is independent of the direction of the observable, but not on the site.
 }
\label{g_Half}
\end{figure}
Next, we study cross-correlations of $\psi$ with different operators $A$, sitting at the L(eft) edge, R(ight) edge, or the M(iddle) site of the chain, 
\begin{equation}
    A=\sigma^{z, x}_L =  \sigma^{z, x}_{site = 1}
    \ , \quad 
    A=\sigma^{z, x}_M =  \sigma^{z, x}_{site = L'+1}
     \ , \quad 
    A=\sigma^{z, x}_R =  \sigma^{z, x}_{site = L}\ .
\end{equation}
We plot $f_A(\omega)g_{\psi,A}(\omega)$ for different $A$ in Fig.~\ref{fig_din_zoom_psiHalf}. 
In the second row of Fig.~\ref{fig_din_zoom_psiHalf}, we display the scaling with the system size of Eqs.\eqref{outofeqETH}, finding a good agreement with the predictions at zero and finite frequency ($\omega=2.1$ shown in the Figure).

Finally, in Fig.~\ref{g_Half} we illustrate  approximate independence of $g_{\psi,A}(\omega)$ at high frequencies,  on the choice of the observable $A$.

\end{appendix}

\bibliography{biblio.bib}

\begin{thebibliography}{10}
\providecommand{\url}[1]{\texttt{#1}}
\providecommand{\urlprefix}{URL }
\expandafter\ifx\csname urlstyle\endcsname\relax
  \providecommand{\doi}[1]{doi:\discretionary{}{}{}#1}\else
  \providecommand{\doi}{doi:\discretionary{}{}{}\begingroup
  \urlstyle{rm}\Url}\fi
\providecommand{\eprint}[2][]{\url{#2}}

\bibitem{deutsch1991quantum}
J.~M. Deutsch,
\newblock \emph{Quantum statistical mechanics in a closed system},
\newblock Physical Review A \textbf{43}(4), 2046 (1991).

\bibitem{srednicki1999approach}
M.~Srednicki,
\newblock \emph{The approach to thermal equilibrium in quantized chaotic
  systems},
\newblock Journal of Physics A: Mathematical and General \textbf{32}(7), 1163
  (1999).

\bibitem{Rigol_2008}
M.~Rigol, V.~Dunjko and M.~Olshanii,
\newblock \emph{Thermalization and its mechanism for generic isolated quantum
  systems},
\newblock Nature \textbf{452}(7189), 854–858 (2008),
\newblock \doi{10.1038/nature06838}.

\bibitem{dalessio2016from}
L.~D'Alessio, Y.~Kafri, A.~Polkovnikov and M.~Rigol,
\newblock \emph{From quantum chaos and eigenstate thermalization to statistical
  mechanics and thermodynamics},
\newblock Advances in Physics \textbf{65}(3), 239 (2016).

\bibitem{khatami2013fluctuation}
E.~Khatami, G.~Pupillo, M.~Srednicki and M.~Rigol,
\newblock \emph{Fluctuation-dissipation theorem in an isolated system of
  quantum dipolar bosons after a quench},
\newblock arXiv preprint arXiv:1304.7279  (2013).

\bibitem{prosen1999}
T.~Prosen,
\newblock \emph{Ergodic properties of a generic nonintegrable quantum many-body
  system in the thermodynamic limit},
\newblock Phys. Rev. E \textbf{60}, 3949 (1999),
\newblock \doi{10.1103/PhysRevE.60.3949}.

\bibitem{rigol2008thermalization}
M.~Rigol, V.~Dunjko and M.~Olshanii,
\newblock \emph{Thermalization and its mechanism for generic isolated quantum
  systems},
\newblock Nature \textbf{452}(7189), 854 (2008).

\bibitem{biroli2010effect}
G.~Biroli, C.~Kollath and A.~M. L{\"a}uchli,
\newblock \emph{Effect of rare fluctuations on the thermalization of isolated
  quantum systems},
\newblock Physical review letters \textbf{105}(25), 250401 (2010),
\newblock \doi{10.1103/PhysRevLett.105.250401}.

\bibitem{ikeda2013finite}
T.~N. Ikeda, Y.~Watanabe and M.~Ueda,
\newblock \emph{Finite-size scaling analysis of the eigenstate thermalization
  hypothesis in a one-dimensional interacting bose gas},
\newblock Physical Review E \textbf{87}(1), 012125 (2013),
\newblock \doi{10.1103/PhysRevE.87.012125}.

\bibitem{steinigeweg2013eigenstate}
R.~Steinigeweg, J.~Herbrych and P.~Prelov{\v{s}}ek,
\newblock \emph{Eigenstate thermalization within isolated spin-chain systems},
\newblock Physical Review E \textbf{87}(1), 012118 (2013),
\newblock \doi{0.1103/PhysRevE.87.012118}.

\bibitem{alba2015eigenstate}
V.~Alba,
\newblock \emph{Eigenstate thermalization hypothesis and integrability in
  quantum spin chains},
\newblock Physical Review B \textbf{91}(15), 155123 (2015),
\newblock \doi{10.1103/PhysRevB.91.155123}.

\bibitem{beugeling2015off}
W.~Beugeling, R.~Moessner and M.~Haque,
\newblock \emph{Off-diagonal matrix elements of local operators in many-body
  quantum systems},
\newblock Physical Review E \textbf{91}(1), 012144 (2015),
\newblock \doi{10.1103/PhysRevE.91.012144}.

\bibitem{luitz2016long}
D.~J. Luitz,
\newblock \emph{Long tail distributions near the many-body localization
  transition},
\newblock Physical Review B \textbf{93}(13), 134201 (2016),
\newblock \doi{10.1103/PhysRevB.93.134201}.

\bibitem{fritzsch2021eigenstate}
F.~Fritzsch and T.~Prosen,
\newblock \emph{Eigenstate thermalization in dual-unitary quantum circuits:
  Asymptotics of spectral functions},
\newblock Phys. Rev. E \textbf{103}, 062133 (2021),
\newblock \doi{10.1103/PhysRevE.103.062133}.

\bibitem{FK2019}
L.~Foini and J.~Kurchan,
\newblock \emph{Eigenstate thermalization hypothesis and out of time order
  correlators},
\newblock Physical Review E \textbf{99}(4), 042139 (2019).

\bibitem{Pappalardi:2022aaz}
S.~Pappalardi, L.~Foini and J.~Kurchan,
\newblock \emph{{Eigenstate Thermalization Hypothesis and Free Probability}},
\newblock Phys. Rev. Lett. \textbf{129}(17), 170603 (2022),
\newblock \doi{10.1103/PhysRevLett.129.170603},
\newblock \eprint{2204.11679}.

\bibitem{pappalardi2023general}
S.~Pappalardi, F.~Fritzsch and T.~Prosen,
\newblock \emph{General eigenstate thermalization via free cumulants in quantum
  lattice systems},
\newblock arXiv preprint arXiv:2303.00713  (2023).

\bibitem{fava2023designs}
M.~Fava, J.~Kurchan and S.~Pappalardi,
\newblock \emph{Designs via free probability},
\newblock arXiv preprint arXiv:2308.06200  (2023).

\bibitem{foini_RotInv}
L.~Foini and J.~Kurchan,
\newblock \emph{Eigenstate thermalization and rotational invariance in ergodic
  quantum systems},
\newblock Phys. Rev. Lett. \textbf{123}, 260601 (2019),
\newblock \doi{10.1103/PhysRevLett.123.260601}.

\bibitem{richter2020eigenstate}
J.~Richter, A.~Dymarsky, R.~Steinigeweg and J.~Gemmer,
\newblock \emph{Eigenstate thermalization hypothesis beyond standard
  indicators: Emergence of random-matrix behavior at small frequencies},
\newblock Physical Review E \textbf{102}(4) (2020).

\bibitem{wang2021eigenstate}
J.~Wang, M.~H. Lamann, J.~Richter, R.~Steinigeweg, A.~Dymarsky and J.~Gemmer,
\newblock \emph{Eigenstate thermalization hypothesis and its deviations from
  random-matrix theory beyond the thermalization time},
\newblock arXiv preprint arXiv:2110.04085  (2021).

\bibitem{brenes2021out}
M.~Brenes, S.~Pappalardi, M.~T. Mitchison, J.~Goold and A.~Silva,
\newblock \emph{Out-of-time-order correlations and the fine structure of
  eigenstate thermalization},
\newblock Physical Review E \textbf{104}(3) (2021).

\bibitem{jafferis2022matrix}
D.~L. Jafferis, D.~K. Kolchmeyer, B.~Mukhametzhanov and J.~Sonner,
\newblock \emph{Matrix models for eigenstate thermalization},
\newblock Physical Review X \textbf{13}(3), 031033 (2023),
\newblock \doi{10.1103/PhysRevX.13.031033}.

\bibitem{jafferis2022jt}
D.~L. Jafferis, D.~K. Kolchmeyer, B.~Mukhametzhanov and J.~Sonner,
\newblock \emph{Jt gravity with matter, generalized eth, and random matrices},
\newblock arXiv preprint arXiv:2209.02131  (2022).

\bibitem{wang2023emergence}
J.~Wang, J.~Richter, M.~H. Lamann, R.~Steinigeweg, J.~Gemmer and A.~Dymarsky,
\newblock \emph{Emergence of unitary symmetry of microcanonically truncated
  operators in chaotic quantum systems},
\newblock arXiv preprint arXiv:2310.20264  (2023).

\bibitem{deBoer2024multiboundary}
J.~de~Boer, D.~Liska and B.~Post,
\newblock \emph{Multiboundary wormholes and ope statistics},
\newblock arXiv preprint arXiv:2405.13111  (2024).

\bibitem{Dymarsky:2016ntg}
A.~Dymarsky, N.~Lashkari and H.~Liu,
\newblock \emph{{Subsystem ETH}},
\newblock Phys. Rev. E \textbf{97}, 012140 (2018),
\newblock \doi{10.1103/PhysRevE.97.012140},
\newblock \eprint{1611.08764}.

\bibitem{chan2019eigenstate}
A.~Chan, A.~De~Luca and J.~T. Chalker,
\newblock \emph{Eigenstate correlations, thermalization, and the butterfly
  effect},
\newblock Phys. Rev. Lett. \textbf{122}, 220601 (2019),
\newblock \doi{10.1103/PhysRevLett.122.220601}.

\bibitem{huang2019universal}
Y.~Huang,
\newblock \emph{Universal eigenstate entanglement of chaotic local
  hamiltonians},
\newblock Nuclear Physics B \textbf{938}, 594 (2019).

\bibitem{PhysRevE.99.032111}
T.-C. Lu and T.~Grover,
\newblock \emph{Renyi entropy of chaotic eigenstates},
\newblock Phys. Rev. E \textbf{99}, 032111 (2019),
\newblock \doi{10.1103/PhysRevE.99.032111}.

\bibitem{murthy2019structure}
C.~Murthy and M.~Srednicki,
\newblock \emph{Structure of chaotic eigenstates and their entanglement
  entropy},
\newblock Physical Review E \textbf{100}(2), 022131 (2019).

\bibitem{shi2023local}
Z.~D. Shi, S.~Vardhan and H.~Liu,
\newblock \emph{Local dynamics and the structure of chaotic eigenstates},
\newblock Phys. Rev. B \textbf{108}, 224305 (2023),
\newblock \doi{10.1103/PhysRevB.108.224305}.

\bibitem{hahn2023statistical}
D.~Hahn, D.~J. Luitz and J.~Chalker,
\newblock \emph{The statistical properties of eigenstates in chaotic many-body
  quantum systems},
\newblock arXiv preprint arXiv:2309.12982  (2023).

\bibitem{jindal2024generalized}
S.~Jindal and P.~Hosur,
\newblock \emph{Generalized free cumulants for quantum chaotic systems},
\newblock arXiv preprint arXiv:2401.13829  (2024).

\bibitem{richter2019impact}
J.~Richter, J.~Gemmer and R.~Steinigeweg,
\newblock \emph{Impact of eigenstate thermalization on the route to
  equilibrium},
\newblock Physical Review E \textbf{99}(5), 050104 (2019).

\bibitem{dymarsky2019mechanism}
A.~Dymarsky,
\newblock \emph{Mechanism of macroscopic equilibration of isolated quantum
  systems},
\newblock Physical Review B \textbf{99}(22), 224302 (2019).

\bibitem{knipschild2020modern}
L.~Knipschild and J.~Gemmer,
\newblock \emph{Modern concepts of quantum equilibration do not rule out
  strange relaxation dynamics},
\newblock Physical Review E \textbf{101}(6), 062205 (2020).

\bibitem{REIMANN2020121840}
P.~Reimann and J.~Gemmer,
\newblock \emph{Why are macroscopic experiments reproducible? imitating the
  behavior of an ensemble by single pure states},
\newblock Physica A: Statistical Mechanics and its Applications \textbf{552},
  121840 (2020),
\newblock \doi{https://doi.org/10.1016/j.physa.2019.121840},
\newblock Tributes of Non-equilibrium Statistical Physics.

\bibitem{dymarsky2022bound}
A.~Dymarsky,
\newblock \emph{Bound on eigenstate thermalization from transport},
\newblock Physical Review Letters \textbf{128}(19), 190601 (2022).

\bibitem{Capizzi:2024msk}
L.~Capizzi, J.~Wang, X.~Xu, L.~Mazza and D.~Poletti,
\newblock \emph{{Hydrodynamics and the eigenstate thermalization hypothesis}}
  (2024),
\newblock \eprint{2405.16975}.

\bibitem{deBoer2024}
J.~de~Boer, D.~Liska, B.~Post and M.~Sasieta,
\newblock \emph{A principle of maximum ignorance for semiclassical gravity},
\newblock Journal of High Energy Physics \textbf{2024}(2), 1 (2024).

\bibitem{essler2016quench}
F.~H. Essler and M.~Fagotti,
\newblock \emph{Quench dynamics and relaxation in isolated integrable quantum
  spin chains},
\newblock Journal of Statistical Mechanics: Theory and Experiment
  \textbf{2016}(6), 064002 (2016),
\newblock \doi{10.1088/1742-5468/2016/06/064002}.

\bibitem{PhysRevLett.122.070601}
I.~M. Khaymovich, M.~Haque and P.~A. McClarty,
\newblock \emph{Eigenstate thermalization, random matrix theory, and
  behemoths},
\newblock Phys. Rev. Lett. \textbf{122}, 070601 (2019),
\newblock \doi{10.1103/PhysRevLett.122.070601}.

\bibitem{mondaini2017}
R.~Mondaini and M.~Rigol,
\newblock \emph{Eigenstate thermalization in the two-dimensional transverse
  field ising model. ii. off-diagonal matrix elements of observables},
\newblock Physical Review E \textbf{96}(1), 012157 (2017).

\bibitem{noh202numerical}
J.~D. Noh, T.~Sagawa and J.~Yeo,
\newblock \emph{Numerical verification of the fluctuation-dissipation theorem
  for isolated quantum systems},
\newblock Phys. Rev. Lett. \textbf{125}, 050603 (2020),
\newblock \doi{10.1103/PhysRevLett.125.050603}.

\bibitem{murthy2019bounds}
C.~Murthy and M.~Srednicki,
\newblock \emph{Bounds on chaos from the eigenstate thermalization hypothesis},
\newblock Physical review letters \textbf{123}(23), 230606 (2019).

\bibitem{berry1977regular}
M.~V. Berry,
\newblock \emph{Regular and irregular semiclassical wavefunctions},
\newblock Journal of Physics A: Mathematical and General \textbf{10}(12), 2083
  (1977),
\newblock \doi{10.1088/0305-4470/10/12/016}.

\bibitem{porter1056fluctuation}
C.~E. Porter and R.~G. Thomas,
\newblock \emph{Fluctuations of nuclear reaction widths},
\newblock Phys. Rev. \textbf{104}, 483 (1956),
\newblock \doi{10.1103/PhysRev.104.483}.

\bibitem{haake1991quantum}
F.~Haake,
\newblock \emph{Quantum signatures of chaos},
\newblock Springer (1991).

\bibitem{luitz2017anomalous}
D.~J. Luitz and Y.~Bar~Lev,
\newblock \emph{Anomalous thermalization in ergodic systems},
\newblock Phys. Rev. Lett. \textbf{117}, 170404 (2016),
\newblock \doi{10.1103/PhysRevLett.117.170404}.

\bibitem{leblond2019entanglement}
T.~LeBlond, K.~Mallayya, L.~Vidmar and M.~Rigol,
\newblock \emph{Entanglement and matrix elements of observables in interacting
  integrable systems},
\newblock Phys. Rev. E \textbf{100}, 062134 (2019),
\newblock \doi{10.1103/PhysRevE.100.062134}.

\bibitem{brenes2020low}
M.~Brenes, J.~Goold and M.~Rigol,
\newblock \emph{Low-frequency behavior of off-diagonal matrix elements in the
  integrable xxz chain and in a locally perturbed quantum-chaotic xxz chain},
\newblock Phys. Rev. B \textbf{102}, 075127 (2020),
\newblock \doi{10.1103/PhysRevB.102.075127}.

\bibitem{foini2022annealed}
L.~Foini and J.~Kurchan,
\newblock \emph{Annealed averages in spin and matrix models},
\newblock SciPost Physics \textbf{12}(3), 080 (2022).

\bibitem{gorin2006dynamics}
T.~Gorin, T.~Prosen, T.~H. Seligman and M.~{\v{Z}}nidari{\v{c}},
\newblock \emph{Dynamics of loschmidt echoes and fidelity decay},
\newblock Physics Reports \textbf{435}(2-5), 33 (2006).

\bibitem{torres2014local}
E.~Torres-Herrera and L.~F. Santos,
\newblock \emph{Local quenches with global effects in interacting quantum
  systems},
\newblock Physical Review E \textbf{89}(6), 062110 (2014).

\bibitem{schiulaz2019thouless}
M.~Schiulaz, E.~J. Torres-Herrera and L.~F. Santos,
\newblock \emph{Thouless and relaxation time scales in many-body quantum
  systems},
\newblock Phys. Rev. B \textbf{99}, 174313 (2019),
\newblock \doi{10.1103/PhysRevB.99.174313}.

\bibitem{rai2023matrix}
K.~S. Rai, J.~I. Cirac and {\'A}.~M. Alhambra,
\newblock \emph{Matrix product state approximations to quantum states of low
  energy variance},
\newblock arXiv preprint arXiv:2307.05200  (2023).

\bibitem{santos2012chaos}
L.~F. Santos, F.~Borgonovi and F.~Izrailev,
\newblock \emph{Chaos and statistical relaxation in quantum systems of
  interacting particles},
\newblock Physical review letters \textbf{108}(9), 094102 (2012).

\bibitem{PhysRevE.85.036209}
L.~F. Santos, F.~Borgonovi and F.~M. Izrailev,
\newblock \emph{Onset of chaos and relaxation in isolated systems of
  interacting spins: Energy shell approach},
\newblock Phys. Rev. E \textbf{85}, 036209 (2012),
\newblock \doi{10.1103/PhysRevE.85.036209}.

\bibitem{micklitz2022emergence}
T.~Micklitz, A.~Morningstar, A.~Altland and D.~A. Huse,
\newblock \emph{Emergence of fermi's golden rule},
\newblock Phys. Rev. Lett. \textbf{129}, 140402 (2022),
\newblock \doi{10.1103/PhysRevLett.129.140402}.

\bibitem{long2023beyond}
D.~M. Long, D.~Hahn, M.~Bukov and A.~Chandran,
\newblock \emph{Beyond fermi's golden rule with the statistical jacobi
  approximation},
\newblock SciPost Physics \textbf{15}(6), 251 (2023),
\newblock \doi{10.21468/SciPostPhys.15.6.251}.

\bibitem{mukerjee2006statistical}
S.~Mukerjee, V.~Oganesyan and D.~Huse,
\newblock \emph{Statistical theory of transport by strongly interacting lattice
  fermions},
\newblock Physical Review B \textbf{73}(3), 035113 (2006),
\newblock \doi{10.1103/PhysRevB.73.035113}.

\bibitem{lux2014hydrodynamic}
J.~Lux, J.~M{\"u}ller, A.~Mitra and A.~Rosch,
\newblock \emph{Hydrodynamic long-time tails after a quantum quench},
\newblock Physical Review A \textbf{89}(5), 053608 (2014).

\bibitem{michailidis2023corrections}
A.~A. Michailidis, D.~A. Abanin and L.~V. Delacr{\'e}taz,
\newblock \emph{Corrections to diffusion in interacting quantum systems},
\newblock arXiv preprint arXiv:2310.10564  (2023),
\newblock \doi{https://doi.org/10.48550/arXiv.2310.10564}.

\bibitem{vidmar2016generalized}
L.~Vidmar and M.~Rigol,
\newblock \emph{Generalized gibbs ensemble in integrable lattice models},
\newblock Journal of Statistical Mechanics: Theory and Experiment
  \textbf{2016}(6), 064007 (2016).

\bibitem{essler2023statistics}
F.~Essler and A.~de~Klerk,
\newblock \emph{Statistics of matrix elements of local operators in integrable
  models},
\newblock arXiv preprint arXiv:2307.12410  (2023).

\bibitem{serbyn2021quantum}
M.~Serbyn, D.~A. Abanin and Z.~Papi{\'c},
\newblock \emph{Quantum many-body scars and weak breaking of ergodicity},
\newblock Nature Physics \textbf{17}(6), 675 (2021).

\bibitem{moudgalya2022quantum}
S.~Moudgalya, B.~A. Bernevig and N.~Regnault,
\newblock \emph{Quantum many-body scars and hilbert space fragmentation: a
  review of exact results},
\newblock Reports on Progress in Physics \textbf{85}(8), 086501 (2022),
\newblock \doi{10.1088/1361-6633/ac73a0}.

\bibitem{chandran2023qmbsreview}
A.~Chandran, T.~Iadecola, V.~Khemani and R.~Moessner,
\newblock \emph{Quantum many-body scars: A quasiparticle perspective},
\newblock Annual Review of Condensed Matter Physics \textbf{14}(1), 443 (2023),
\newblock \doi{10.1146/annurev-conmatphys-031620-101617}.

\bibitem{gotta2023asymptotic}
L.~Gotta, S.~Moudgalya and L.~Mazza,
\newblock \emph{Asymptotic quantum many-body scars},
\newblock Phys. Rev. Lett. \textbf{131}, 190401 (2023),
\newblock \doi{10.1103/PhysRevLett.131.190401}.

\bibitem{High_T}
A.~Maillard, L.~Foini, A.~L. Castellanos, F.~Krzakala, M.~M{\'e}zard and
  L.~Zdeborov{\'a},
\newblock \emph{High-temperature expansions and message passing algorithms},
\newblock Journal of Statistical Mechanics: Theory and Experiment
  \textbf{2019}(11), 113301 (2019).

\bibitem{avdoshkin2020euclidean}
A.~Avdoshkin and A.~Dymarsky,
\newblock \emph{Euclidean operator growth and quantum chaos},
\newblock Physical Review Research \textbf{2}(4), 043234 (2020).

\bibitem{Deutsch_2010}
J.~M. Deutsch,
\newblock \emph{Thermodynamic entropy of a many-body energy eigenstate},
\newblock New Journal of Physics \textbf{12}(7), 075021 (2010),
\newblock \doi{10.1088/1367-2630/12/7/075021}.

\end{thebibliography}

\end{document}